\documentclass[pdflatex,sn-mathphys-num]{sn-jnl}

\usepackage{graphicx}%
\usepackage{multirow}%
\usepackage{amsmath,amssymb,amsfonts}%
\usepackage{amsthm}%
\usepackage{mathrsfs}%
\usepackage[title]{appendix}%
\usepackage{xcolor}%
\usepackage{textcomp}%
\usepackage{manyfoot}%
\usepackage{booktabs}%
\usepackage{algorithm}%
\usepackage{algorithmicx}%
\usepackage{algpseudocode}%
\usepackage{listings}%
\usepackage{rotating}
\usepackage{makecell}  

\theoremstyle{thmstyleone}%
\theoremstyle{thmstyletwo}%

\theoremstyle{thmstylethree}%

\begin{document}

\title[Performance Characterization of Distributed Deep Learning Strategies]{Performance Characterization of Distributed Deep Learning Strategies: A Quantitative Evaluation of DDP, FSDP, and Parameter Server Architectures on GPU Clusters}








\author{\fnm{Md Sultanul Islam} \sur{Ovi}}\email{movi@gmu.edu}

\affil{\orgdiv{Department of Computer Science}, \orgname{George Mason University}, \orgaddress{\street{4400 University Dr}, \city{Fairfax}, \postcode{22030}, \state{VA}, \country{USA}}}

\abstract{Efficiently scaling deep neural networks across GPU clusters requires navigating complex trade-offs between computational throughput, memory utilization, and synchronization overhead. This paper presents a unified empirical evaluation of three dominant distributed training paradigms: Distributed Data Parallel (DDP), Fully Sharded Data Parallel (FSDP), and the Parameter Server (PS) architecture. We conduct side-by-side benchmarking on both high-performance (NVIDIA A100) and commodity-class (NVIDIA A10G) clusters to isolate the impact of communication bandwidth and gang-scheduling dependencies. Our results indicate that while DDP achieves a $2\text{--}3\times$ speedup in training throughput for standard architectures, FSDP demonstrates a $4\text{--}6\times$ reduction in peak memory usage, validating its utility for memory-constrained environments despite higher communication latency. Furthermore, we evaluate the elasticity of the Parameter Server architecture; while Asynchronous PS reduced training time by up to 28\% compared to synchronous approaches, it incurred significant accuracy penalties (ranging from 4\% to 17\%) due to gradient staleness. We also analyze a modified, staleness-mitigating asynchronous protocol, which we found introduced synchronization overheads that negated throughput gains. These findings provide a decision framework for system designers, highlighting that while DDP remains optimal for homogeneous, gang-scheduled clusters, FSDP and PS offer critical alternatives for memory-bound and heterogeneous environments respectively.}




\maketitle

\begin{center}
\small
This manuscript is a preprint and is currently under review for journal publication.
\end{center}


\section{Introduction}\label{sec1}

Deep neural networks (DNNs) have grown dramatically in scale, now reaching hundreds of billions or even trillions of parameters in cutting-edge models \cite{dean2012large}. For example, the GPT-3 language model, with 175 billion parameters, demonstrated unprecedented capabilities in natural language processing \cite{mann2020language}, and industry-scale recommendation systems now surpass 1 trillion parameters by employing massive embedding tables \cite{rajbhandari2020zero}. Training these ultra-large models to achieve state-of-the-art accuracy requires distributed computing across large GPU clusters. The current trend involves using distributed training frameworks that coordinate dozens to thousands of accelerators in parallel, reducing training time for large datasets and models \cite{xu2021gspmd}. Recent work shows that data-parallel strategies, model sharding, and mixed parallelism are essential to scaling model size further \cite{zheng2022alpa, zhang2022mics}.

Despite the success of distributed deep learning, challenges related to scalability and efficiency continue to arise. Synchronous data-parallel training, commonly implemented through PyTorch’s Distributed Data Parallel (DDP) module, replicates the model on each GPU and synchronizes gradients using high-speed all-reduce collectives such as NCCL or Gloo \cite{sergeev2018horovod}. DDP has become a standard for multi-GPU training and often achieves near-linear scaling on many vision and NLP tasks \cite{aach2023large}. However, with increasing cluster and model sizes, the cost of gradient synchronization grows, creating communication overhead that can dominate training time \cite{bai2021gradient}. Even with optimized libraries, exchanging gradients may consume 60 to 80 percent of iteration time during transformer model training on 128 GPUs connected via a 100 Gbps network \cite{shi2014designing}. This bottleneck has motivated research into communication-efficient strategies such as gradient compression, which reduces bandwidth usage by sparsifying or quantizing gradients \cite{agarwal2022utility}. Although these methods can reduce gradient size with minimal loss in convergence \cite{tang2020communication}, integrating them into frameworks without affecting performance remains a research challenge \cite{yuan2021oneflow}.

To address memory constraints, model-parallel and sharded training strategies have gained traction. One notable solution is Fully Sharded Data Parallel (FSDP), which extends data-parallelism by partitioning model parameters, gradients, and optimizer states across GPUs \cite{zhao2023pytorch}. This technique, also known as zero redundancy optimization, avoids the memory overhead caused by replicating models on each GPU, enabling training of significantly larger models \cite{rajbhandari2020zero}. PyTorch’s implementations of FSDP, such as FairScale and native support in PyTorch 2.x, include memory-efficient tensor structures and overlapped communication and computation for improved throughput \cite{b28}. Meta AI’s FSDP shows throughput similar to DDP while training models that exceed memory limits of standard data-parallelism \cite{zhao2023pytorch}. Similarly, Microsoft’s ZeRO techniques partition optimizer states and gradients, with ZeRO-3 fully sharding all model states and supporting on-demand parameter gathering \cite{rajbhandari2020zero, rajbhandari2021zero}. ZeRO enabled training trillion-parameter models by removing memory redundancy \cite{rajbhandari2020zero}, and extensions like ZeRO-Offload and ZeRO-Infinity utilize CPU and NVMe memory to bypass GPU memory ceilings \cite{ren2021zero, aach2023large}. Although sharded approaches offer memory advantages, they introduce different communication patterns and runtime behaviors compared to DDP, warranting empirical comparison.

The parameter server (PS) architecture presents an alternative paradigm. In this setup, server nodes maintain global model parameters while worker nodes compute gradients and push updates to the servers \cite{li2014communication}. This architecture allows asynchronous training, where workers operate without strict synchronization, which can improve utilization in heterogeneous clusters. Techniques to mitigate drawbacks such as stale gradients include bounded staleness (as in SSP algorithms) and adaptive learning rate corrections \cite{ho2013more}. Recent systems revisit PS to improve GPU-cluster scalability. BytePS, for example, combines PS and all-reduce by leveraging idle CPU cores for aggregation, improving throughput in mixed hardware clusters \cite{jiang2020unified}. HeterPS adopts hybrid scheduling with reinforcement learning to assign model layers to CPU or GPU, achieving significant throughput and cost gains \cite{liu2023heterps}. These designs show that parameter servers, especially when combined with intelligent scheduling and partial asynchrony, remain relevant for training large models in real-world environments \cite{thangakrishnan2020herring}.

\subsection*{Motivation and Problem Statement}

A major gap in the existing literature is the lack of unified, head-to-head evaluations of these distributed strategies on modern hardware that explicitly account for operational constraints like \textit{gang scheduling}. Most existing research focuses on optimizing a single technique in isolation \cite{ben2019demystifying}. However, in multi-tenant clusters, the strict synchronization required by DDP and FSDP imposes a ``gang scheduling'' constraint, where all GPUs must be available simultaneously and operate in lockstep. This creates resource fragmentation and sensitivity to stragglers. While Parameter Server (PS) architectures offer asynchronous capabilities that relax these scheduling constraints, their trade-offs in terms of convergence accuracy and modern hardware utilization remain under-explored in the context of recent deep learning workloads.

Practitioners currently lack a clear decision framework that balances these conflicting objectives: training throughput (speed), memory efficiency (scalability), and scheduling elasticity (asynchrony). This work addresses that gap through a structured empirical comparison.

\subsection*{Our contributions are as follows:}

\textit{Systematic Comparative Study:} We present a unified empirical evaluation of PyTorch DDP, Fully Sharded Data Parallel (FSDP), and both Synchronous and Asynchronous Parameter Server architectures. We benchmark these strategies on consistent hardware to isolate the overheads of communication versus computation.

\textit{Quantitative Performance Analysis:} We report detailed measurements of training throughput, memory utilization, and convergence behavior. Our results quantify distinct performance profiles: DDP achieved a $\mathbf{2\text{--}3\times}$ speedup in training throughput compared to FSDP on standard clusters, while FSDP demonstrated a $\mathbf{4\text{--}6\times}$ reduction in peak memory usage, validating its utility for memory-constrained environments despite higher communication latency.

\textit{Evaluation of Asynchronous Elasticity:} We critically evaluate the viability of relaxing gang scheduling via Asynchronous PS. We observe that while Asynchronous PS reduces training time by up to 28\% by removing synchronization barriers, it incurs an accuracy penalty of 4\% to 17\% due to gradient staleness. We further demonstrate that naive attempts to mitigate this staleness (Modified Async PS) introduce synchronization overheads that negate the throughput benefits.

\textit{Insights and Best Practices:} Based on our experimental results, we provide a decision matrix for system designers. We identify that DDP is optimal for homogeneous, gang-scheduled clusters; FSDP is essential for maximizing model size on limited hardware; and Asynchronous PS remains a specialized tool for heterogeneous environments where fault tolerance outweighs peak accuracy.

\section{Literature Review}\label{sec2}

Distributed training of deep neural networks has been widely investigated, producing an extensive body of work across multiple parallelization strategies. In this section, we organize prior contributions based on the primary method used: data parallelism, model parallelism, pipeline parallelism, hybrid techniques, parameter server designs, sharded data parallelism, and federated learning. For each category, we summarize its evolution, technical advancements, and known trade-offs between scalability, efficiency, and implementation complexity. Additionally, we review related work on cluster scheduling and communication bottlenecks that impact the practicality of these strategies in shared computing environments.

In the data parallelism domain, early experiments such as AlexNet established that training convolutional neural networks could be substantially accelerated by splitting data batches across multiple GPUs \cite{krizhevsky2012imagenet}. Subsequent work optimized communication protocols to support this strategy at scale, with systems like Horovod and PyTorch DDP enabling near-linear throughput scaling up to hundreds of GPUs \cite{sergeev2018horovod, andersen2014scaling, ben2019demystifying}. To mitigate communication overheads, researchers introduced gradient compression, overlapping techniques, and high-performance collectives such as NCCL \cite{bai2021gradient, agarwal2022utility, tang2020communication, chetlur2014cudnn}.

Model parallelism emerged to address memory limitations by partitioning neural network layers or tensors across devices \cite{shoeybi2019megatron, xu2021gspmd}. This approach allows extremely large models to be trained by distributing storage and compute responsibilities, albeit at the cost of increased inter-device communication. Innovations such as Mixture-of-Experts (MoE) and intra-layer parallelism, demonstrated by GShard and Megatron-LM respectively, showed the feasibility of training models with hundreds of billions of parameters by leveraging sparsity and splitting wide layers \cite{lepikhin2020gshard, rajbhandari2022deepspeed, shazeer2017outrageously}.

Pipeline parallelism was introduced to improve utilization in model-parallel training by streaming micro-batches through different model stages concurrently \cite{huang2019gpipe}. This method reduces idle time and increases throughput, although it introduces complexities like pipeline flushing and weight staleness. Systems such as GPipe and PipeDream developed mechanisms to manage these issues and demonstrated performance improvements on deep transformer models \cite{huang2019gpipe, narayanan2019pipedream}.

Hybrid parallelism has become increasingly prevalent, particularly in large-scale model training. It combines multiple strategies: such as model parallelism for memory constraints, pipeline parallelism for depth, and data parallelism for replication: to achieve both scalability and efficiency \cite{aach2023large, narayanan2021efficient, jia2019beyond}. Projects like DeepSpeed and Megatron-LM have implemented 3D parallelism, which includes tensor, pipeline, and data-parallel components, to train models beyond 100 billion parameters while maintaining practical training speed \cite{rajbhandari2022deepspeed, narayanan2021efficient,akintoye2022hybrid}.

Parameter server (PS) architectures offered early solutions for distributing models and updates asynchronously \cite{dean2012large, ho2013more, li2014communication}. Systems like DistBelief and Petuum demonstrated scalable training across commodity CPU clusters, introducing techniques like bounded staleness and sparse updates. While all-reduce methods have supplanted PS in many GPU-rich environments due to hardware support and simplicity, PS remains relevant for sparse models and federated environments where flexibility and elasticity are required \cite{mcmahan2017communication, li2020federated}.

Fully Sharded Data Parallelism (FSDP) evolved from ZeRO-style optimizers and combines the memory advantages of model parallelism with the synchronization simplicity of data parallelism \cite{zhao2023pytorch, rajbhandari2020zero}. FSDP shards parameters, gradients, and optimizer states across GPUs and reconstructs them only when needed. This significantly reduces per-GPU memory consumption and enables training of trillion-parameter models on modern clusters \cite{zhao2023pytorch, ren2021zero}. Although it introduces communication overhead, it supports near-linear scaling when overlapped correctly with computation.

Federated learning expands the concept of distributed training beyond centralized clusters to privacy-preserving, client-device setups. FL strategies such as FedAvg perform local updates on edge devices followed by global aggregation, ensuring that raw data remains decentralized \cite{mcmahan2017communication}. These methods address bandwidth and heterogeneity challenges, often employing quantized updates, secure aggregation, and personalized model adaptations. Ideas from FL have influenced communication-efficient strategies in GPU cluster training as well \cite{li2020federated, chen2014big}.

Complementary to these training techniques are cluster scheduling and communication system designs. Gang scheduling, elastic job management, and in-network aggregation mechanisms directly affect how distributed training performs in practice, especially in multi-tenant clusters \cite{jiang2020unified, barham2022pathways, aach2023large}. Studies show that scheduling delays, resource fragmentation, and network contention can limit achievable throughput, even when parallel strategies are theoretically sound. Research in this space has proposed dynamic scaling, time-slicing, and gradient-aware scheduling to better utilize hardware \cite{tang2020communication, xu2020automatic, liu2023heterps}.

\begin{sidewaystable}
\caption{Recent Works on Distributed Training Strategies}
\label{tab:dist_lit_summary}
\begin{tabular*}{\textheight}{@{\extracolsep{\fill}}p{3.5cm}p{1.6cm}p{2.2cm}p{5.6cm}p{4cm}}
\toprule
\textbf{Paper Title (Authors)} & \textbf{Venue (Year)} & \textbf{Strategy} & \textbf{Key Contributions} & \textbf{Hardware/Software Used} \\
\midrule
DeepSpeed: System Optimizations... (Rasley et al.) \cite{rasley2020deepspeed} & KDD (2020) & ZeRO (DP) & Introduced ZeRO optimizer reducing memory redundancy, enabling 100B-param model training with 15 PFLOPS throughput. & Up to 400 NVIDIA V100 GPUs (Azure), PyTorch + DeepSpeed. \\
\midrule
ZeRO: Memory Optimizations... (Rajbhandari et al.) \cite{rajbhandari2020zero} & SC (2020) & ZeRO (DP) & Sharded model states across GPUs, scaling to trillion-param models. Achieved 15 PFLOPS on 400 GPUs. & 400 NVIDIA V100 GPUs, DeepSpeed (PyTorch). \\
\midrule
ZeRO-Offload (Ren et al.) \cite{ren2021zero} & USENIX ATC (2021) & ZeRO + CPU Offload & Offloaded optimizer states to CPU memory, enabling 13B-param training on single GPU, scaling to 128 GPUs. & V100 GPUs, CPU RAM, PyTorch + DeepSpeed. \\
\midrule
ZeRO-Infinity (Rajbhandari et al.) \cite{rajbhandari2021zero} & SC (2021) & ZeRO + CPU/NVMe & Used CPU+NVMe to scale model training beyond GPU limits; enabled training 1T-param model on 16 GPUs. & A100 GPUs with 28TB NVMe, PyTorch + DeepSpeed. \\
\midrule
PyTorch FSDP (Zhao et al.) \cite{zhao2023pytorch} & PVLDB (2023) & FSDP (DP) & Introduced scalable sharded training via PyTorch-native FSDP, reducing memory per GPU while matching DDP speed. & Up to 512 A100/V100 GPUs, PyTorch FSDP. \\
\midrule
PatrickStar (Fang et al.) \cite{fang2022parallel} & TPDS (2023) & Hybrid (ZeRO-Offload) & Chunk-based dynamic memory manager enabling 12B GPT model training on 8 GPUs. & 8x V100 GPUs + 240GB RAM, PyTorch + ZeRO. \\
\midrule
TensorOpt (Cai et al.) \cite{cai2021tensoropt} & TPDS (2022) & Auto-Parallel (DP/MP) & Auto-tunes hybrid parallelism under memory/time constraints, supporting CNN and Transformer models. & PyTorch, up to 16 GPUs, ResNet, BERT. \\
\midrule
BytePS (Jiang et al.) \cite{jiang2020unified} & OSDI (2020) & All-Reduce + PS & Unified architecture for PS and all-reduce with CPU-side gradient aggregation, outperforming NCCL and standard PS. & 256-GPU cluster, 100 Gbps IB/Ethernet, TensorFlow/PyTorch/MXNet. \\
\midrule
Megatron-LM (Narayanan et al.) \cite{narayanan2021efficient} & SC (2021) & Hybrid (DP+MP+PP) & Combined 3D parallelism to train 1T-param model using 3072 GPUs with 502 PFLOP/s throughput. & 3072 A100 GPUs, NVLink+IB, PyTorch + Megatron-LM. \\
\midrule
Gradient Compression Study (Agarwal et al.) \cite{agarwal2022utility} & MLSys (2022) & Comm. Compression & Benchmarked 200+ compression setups, finding limited gains in well-optimized data center environments. & 16--32 GPU clusters, PyTorch, Horovod/NCCL. \\
\midrule
TeraPipe (Li et al.) \cite{li2021terapipe} & ICML (2021) & Pipeline (PP) & Proposed token-based pipelining for Transformers, 5× faster GPT-3 training on 384 GPUs vs conventional split. & 384 V100 GPUs (AWS), Megatron-LM + PyTorch. \\
\midrule
Pathways (Barham et al.) \cite{barham2022pathways} & MLSys (2022) & Async (Custom) & Google’s async runtime for cross-pod scheduling, reached 100\% TPU utilization on 2048 cores. & TPUv4 pods, Google JAX + XLA. \\
\botrule
\end{tabular*}
\footnotetext{Note: Paper titles are abbreviated for readability. Full citation details are included in the References section.}
\end{sidewaystable}

In summary, the literature on distributed training reveals a continuous progression from early synchronous strategies to more adaptive, memory-efficient, and communication-aware systems. Techniques like FSDP and hybrid parallelism mark the current state of the art, enabling unprecedented model sizes and training efficiency. Parameter server designs and federated learning extend this capability to new deployment environments, while advancements in scheduling and system-level coordination ensure better utilization of shared compute resources.

Our work builds upon these foundations by empirically comparing DDP, FSDP, and Parameter Server strategies on standardized workloads and hardware. Through a controlled benchmarking setup, we aim to clarify their practical trade-offs and identify best-fit scenarios for each strategy. These findings are intended to inform future system designers and practitioners seeking to deploy large-scale deep learning workloads efficiently.

Table~\ref{tab:dist_lit_summary} provides a concise overview of twelve influential papers in distributed deep learning, categorized by their primary training strategy. For each work, we list the title, authors, publication venue and year, the main method used (DDP, FSDP, ZeRO, PS), its technical contributions, and the hardware/software platforms used in evaluation.

\section{Research Methodology}\label{sec:methodology}

\begin{figure}[ht]
\centering
    \makebox[\linewidth][c]{%
        \includegraphics[width=1.3\linewidth]{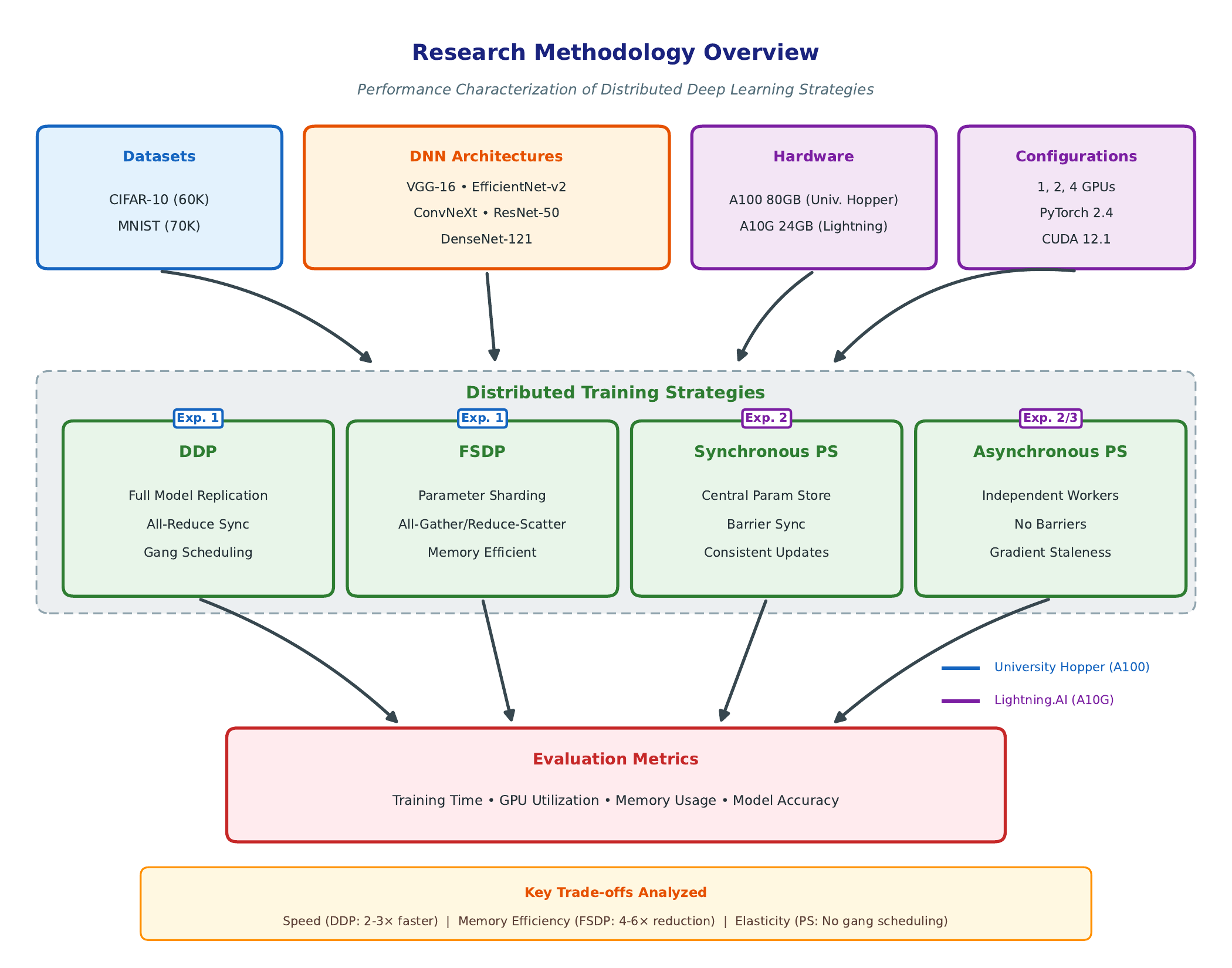}
    }
\caption{Research methodology overview illustrating the experimental framework for evaluating distributed training strategies. The study examines DDP, FSDP, and Parameter Server architectures across two GPU cluster environments (University Hopper with A100 GPUs and Lightning.AI with A10G GPUs) using multiple DNN architectures and evaluation metrics.}\label{fig:methodology_overview}
\end{figure}

This research addresses a critical limitation in contemporary distributed deep learning systems: the dependence on gang scheduling paradigms that require synchronized resource allocation across multiple compute nodes. Traditional distributed training approaches rely on synchronized scheduling mechanisms to coordinate tasks across multiple Graphics Processing Units (GPUs), which can impose significant constraints on scalability and operational efficiency \cite{jiang2020unified}. This study investigates various distributed deep learning strategies to evaluate their potential for reducing or eliminating dependence on gang scheduling requirements.

The primary objective of this investigation is to conduct a systematic analysis of the effectiveness of three distinct distributed training strategies: Data Parallelism via Distributed Data Parallel (DDP), Fully Sharded Data Parallelism (FSDP), and the Parameter Server (PS) approach. These strategies are evaluated across multiple deep learning model architectures to assess their impact on GPU utilization efficiency, memory consumption patterns, model accuracy, and training duration. Through controlled evaluation under varying computational loads and network conditions, this study seeks to identify methodologies that optimally enhance scalability and efficiency of model training without the constraints imposed by synchronized collective operations.

\subsection{Study Design and Rationale}
To rigorously evaluate the overheads of distributed strategies, we prioritized a \textit{communication-bound} experimental design. We selected standard datasets (CIFAR-10, MNIST) and diverse model architectures to create a scenario where the computation-to-communication ratio is relatively low. This design choice stresses the synchronization primitives (All-Reduce, All-Gather) and exposes the latency costs of each strategy, which are often masked in compute-bound, large-batch training scenarios. This approach allows us to isolate the specific efficiency bottlenecks of DDP, FSDP, and PS architectures. Figure~\ref{fig:methodology_overview} provides a visual overview of our experimental framework, illustrating the relationships between datasets, model architectures, distributed training strategies, and evaluation metrics.

\subsection{Dataset Description}

\textbf{CIFAR-10 Dataset: } The CIFAR-10 dataset \cite{krizhevsky2012imagenet} constitutes a labeled subset of the comprehensive CIFAR collection. While relatively small, it serves as an effective proxy for establishing baseline communication overheads in distributed vision tasks. The dataset comprises 60,000 color images with dimensions of 32$\times$32 pixels distributed across ten distinct classes, with 6,000 images per class. The classification categories include airplanes, automobiles, birds, cats, deer, dogs, frogs, horses, ships, and trucks.

\begin{itemize}
    \item \textbf{Number of Classes:} 10
    \item \textbf{Image Dimensions:} 32$\times$32 pixels
    \item \textbf{Total Images:} 60,000
\end{itemize}

\textbf{MNIST Dataset: } The Modified National Institute of Standards and Technology (MNIST) dataset represents a comprehensive database of handwritten digits commonly employed for benchmarking. We utilize MNIST to evaluate the Parameter Server architecture in bandwidth-constrained environments, providing a granular view of gradient staleness effects. The MNIST collection contains 70,000 grayscale images of handwritten digits, each with dimensions of 28$\times$28 pixels, distributed across ten classes representing digits 0 through 9.

\begin{itemize}
    \item \textbf{Number of Classes:} 10
    \item \textbf{Image Dimensions:} 28$\times$28 pixels
    \item \textbf{Total Images:} 70,000
\end{itemize}

\subsection{Deep Learning Model Architectures}
We selected a diverse set of models to evaluate different resource bottlenecks:

\textbf{ConvNeXt-Large Model: } The ConvNeXt-Large architecture \cite{liu2022convnet} represents an advanced convolutional neural network design. This model (approx. 198M parameters) was selected to represent a \textbf{compute-intensive} workload, testing the ability of distributed strategies to overlap heavy computation with communication.

\textbf{VGG-16 Model: } The Visual Geometry Group 16-layer network (VGG-16) \cite{simonyan2014very}, comprising 138.4 million parameters, is characterized by its deep yet architecturally streamlined layer configuration. We included VGG-16 as a \textbf{parameter-heavy} baseline to stress-test memory consumption and parameter transfer bandwidth, particularly for the FSDP and PS evaluations.

\textbf{EfficientNet-v2 Model: } The EfficientNet-v2 architecture \cite{tan2021efficientnetv2} builds upon its predecessor to deliver enhanced accuracy while maintaining computational efficiency. With 118.5 million parameters, it serves as a balanced workload to benchmark optimization efficiency.

\textbf{ResNet-50 Model: } The Residual Network with 50 layers (ResNet-50) \cite{he2016deep}, containing approximately 25.6 million parameters, addresses the fundamental challenge of training deep neural networks. It is used here as a standard industry baseline for scaling efficiency.

\textbf{DenseNet-121 Model: } The Densely Connected Convolutional Network with 121 layers (DenseNet-121) \cite{huang2017densely}, comprising 8 million parameters, employs a distinctive dense connectivity pattern. This model tests the handling of complex dependency graphs in distributed backward passes.

\subsection{Distributed Training Strategies}

\begin{figure}[ht]
\centering
\includegraphics[width=0.9\textwidth]{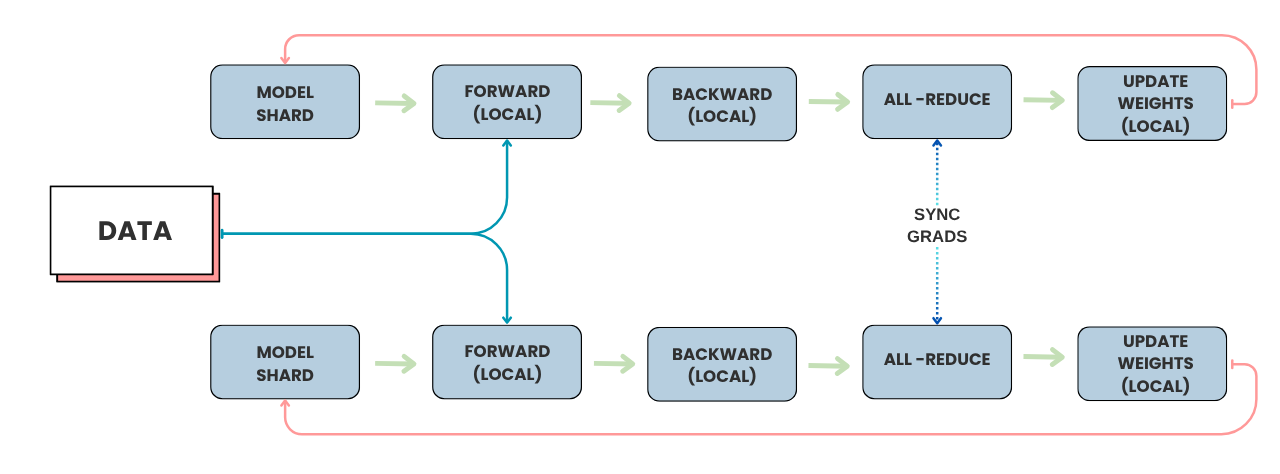}
\caption{Distributed Data Parallel (DDP) strategy where the full model is replicated on each GPU and gradients are synchronized using all-reduce.}\label{fig:ddp_overview}
\end{figure}

\textbf{Distributed Data Parallel (DDP): } Distributed Data Parallel represents a parallelization strategy designed to leverage the computational capabilities of multiple Graphics Processing Units (GPUs) through complete model replication across each computational unit. This methodology \cite{sergeev2018horovod} enables parallel processing of distinct data partitions during both forward and backward propagation phases. The gradients computed by individual GPUs are subsequently synchronized through an all-reduce collective operation, resulting in a unified model parameter update across all participating devices.

While this approach optimizes parallel computational throughput, it necessitates substantial memory overhead due to the complete replication of model parameters, gradients, and optimizer states on each GPU. Figure~\ref{fig:ddp_overview} illustrates the DDP architecture, demonstrating how each GPU maintains a complete copy of the model while processing different data batches. The all-reduce operation ensures gradient consistency across all replicas before parameter updates are applied uniformly \cite{bai2021gradient}.

\begin{figure}[ht]
\centering
\includegraphics[width=\textwidth]{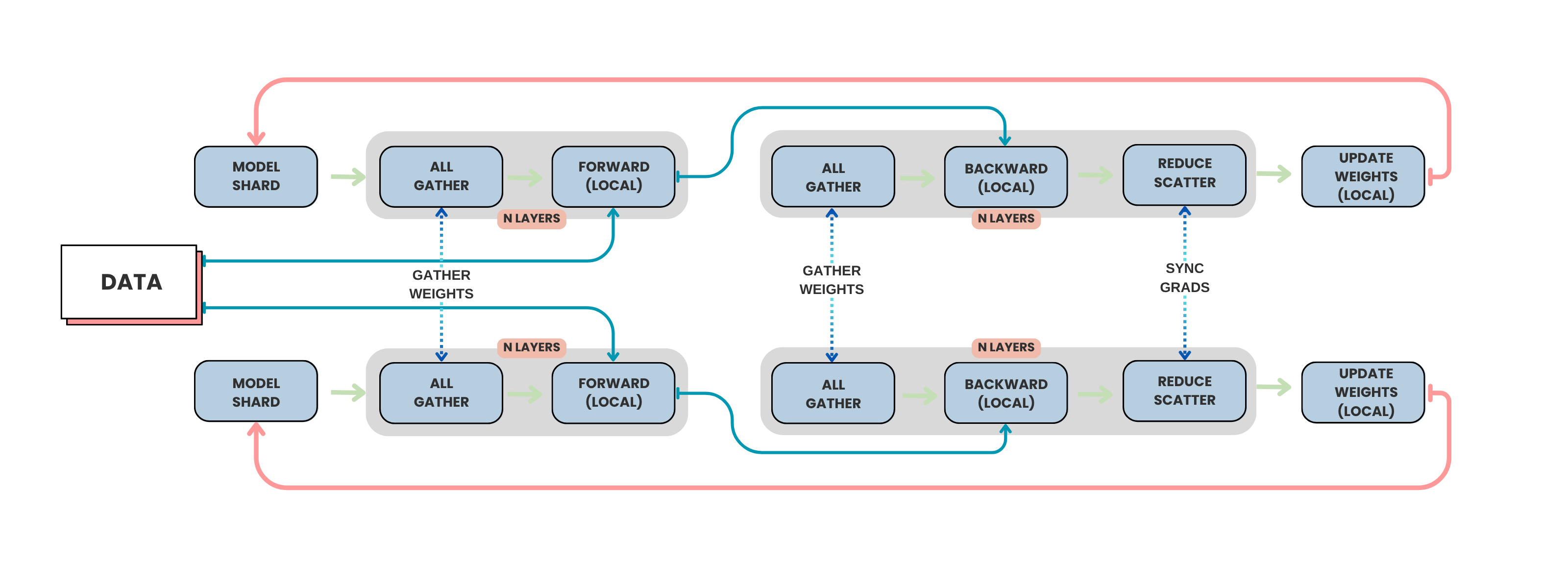}
\caption{Fully Sharded Data Parallel (FSDP) strategy with model parameters, gradients, and optimizer states partitioned across GPUs.}\label{fig:fsdp_overview}
\end{figure}

\textbf{Fully Sharded Data Parallel (FSDP): } Fully Sharded Data Parallel introduces an alternative paradigm wherein model parameters are partitioned, or sharded, across participating GPUs \cite{zhao2023pytorch}. Each GPU maintains only a subset of the complete model parameters, thereby significantly reducing per-GPU memory requirements. During the forward propagation phase, the strategy employs an all-gather collective operation to retrieve necessary parameters from all participating GPUs for computation.

The backward propagation follows a complementary pattern, utilizing a reduce-scatter operation to efficiently aggregate and distribute gradients for localized parameter updates. This approach, depicted in Figure~\ref{fig:fsdp_overview}, enhances scalability by enabling training of substantially larger models through effective memory resource management. The figure demonstrates how parameters are distributed across GPUs and dynamically gathered during computation phases, allowing memory-efficient training of models that exceed individual GPU memory capacity \cite{zhao2023pytorch}.

\begin{figure}[ht]
\centering
\includegraphics[width=0.7\textwidth]{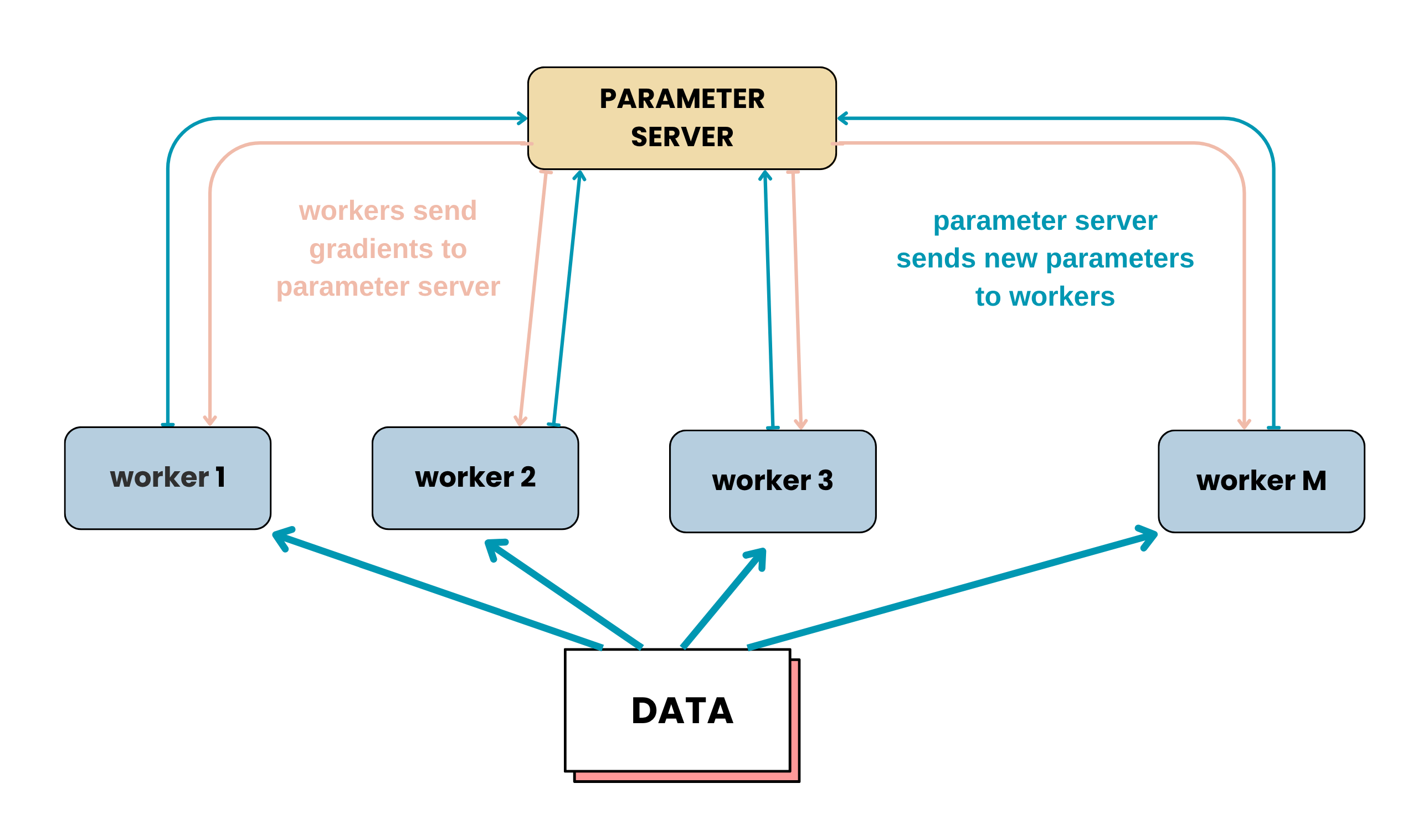}
\caption{Parameter Server architecture where model parameters are centrally maintained and accessed by distributed worker nodes.}\label{fig:ps}
\end{figure}

\begin{figure}[ht]
\centering
\includegraphics[width=0.7\textwidth]{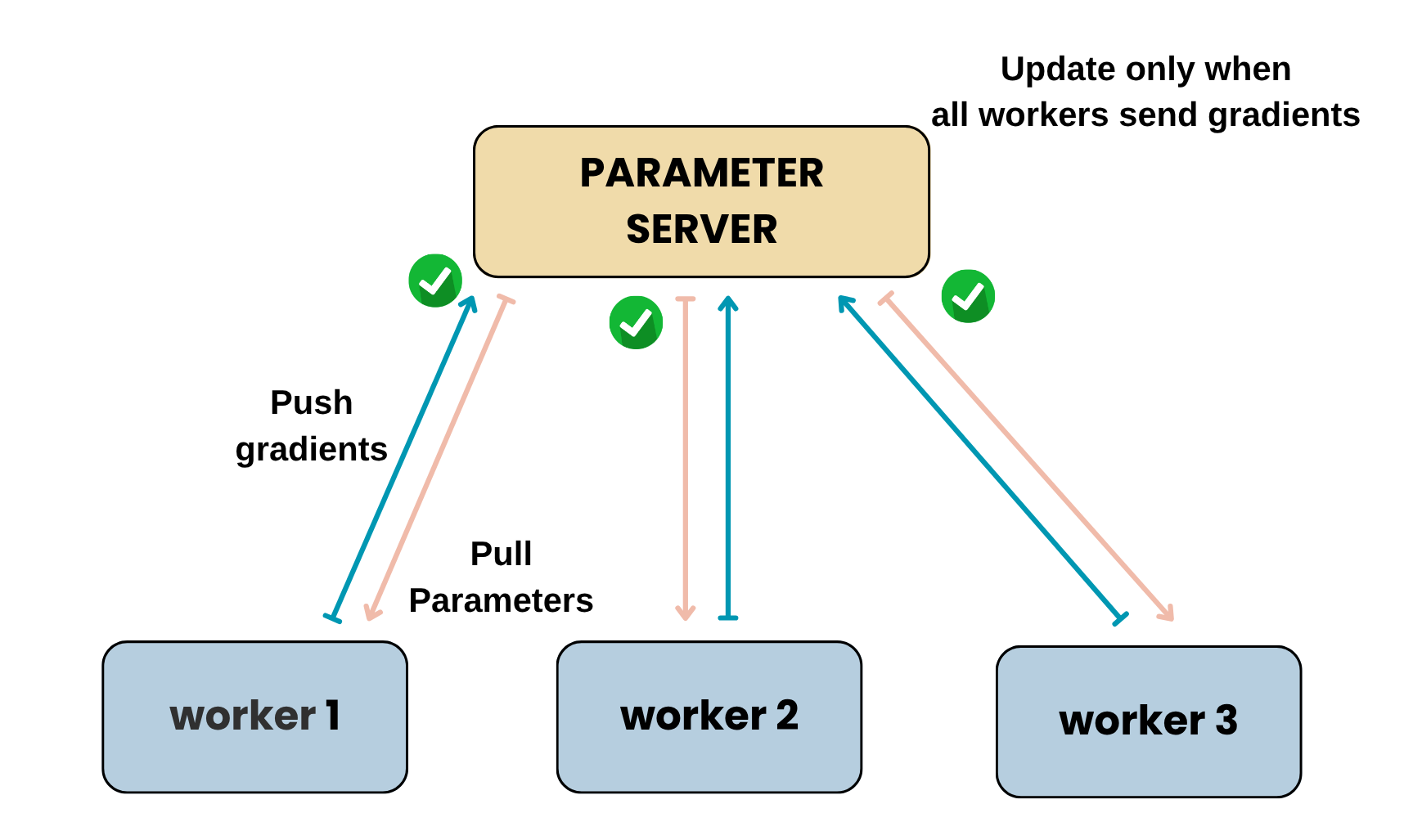}
\caption{Synchronous Parameter Server: Workers compute gradients in parallel and update parameters synchronously.}\label{fig:pss}
\end{figure}

\begin{figure}[ht]
\centering
\includegraphics[width=\textwidth]{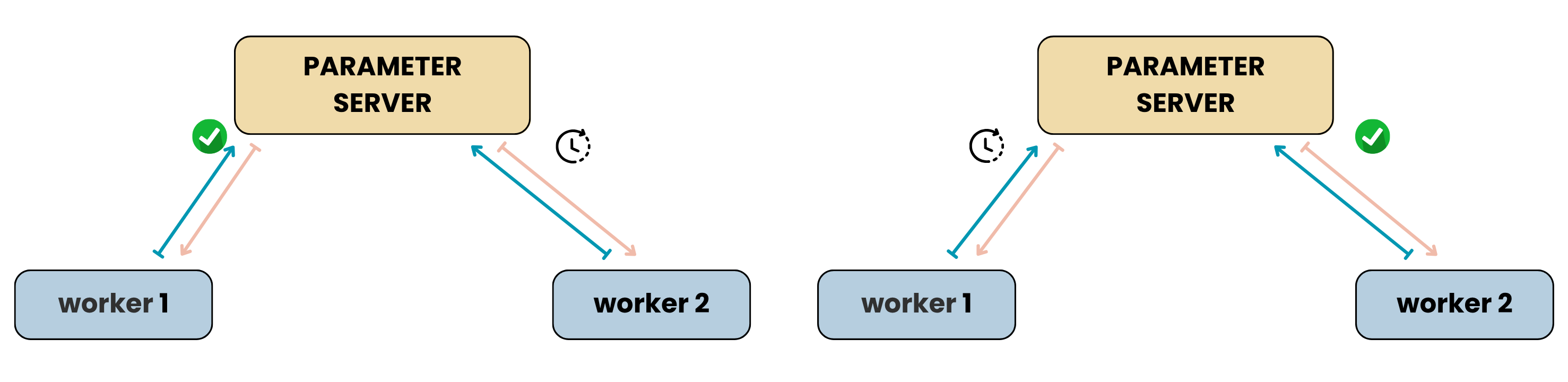}
\caption{Asynchronous Parameter Server: Workers update parameters independently, reducing idle time but risking inconsistency.}\label{fig:psa}
\end{figure}

\textbf{Parameter Server Architecture: } The Parameter Server (PS) model \cite{li2014communication} implements a decentralized training paradigm that separates the computational workload of gradient computation from the storage and management of model parameters. Within this architectural framework, worker nodes assume responsibility for computing gradients using their allocated data partitions, while dedicated parameter server nodes maintain and manage model parameters centrally.

Figure~\ref{fig:ps} illustrates the fundamental Parameter Server architecture, emphasizing the central role of parameter servers in maintaining model state across the distributed system. Worker nodes communicate with parameter servers to retrieve current parameters and submit computed gradient updates during training iterations. This separation of concerns enables flexible resource allocation and can accommodate heterogeneous computational environments. The Parameter Server approach can be implemented in two distinct operational modes:

\textbf{Synchronous Parameter Server:} In the synchronous configuration, all worker nodes compute gradients in a coordinated manner and update parameters through synchronized operations. Figure~\ref{fig:pss} demonstrates this synchronous update mechanism, which ensures consistency of model parameters across all workers at the expense of computational efficiency, as all nodes must await completion of the slowest worker before proceeding to subsequent training iterations.

\textbf{Asynchronous Parameter Server:} The asynchronous variant permits worker nodes to independently compute and apply parameter updates without synchronization barriers with other computational nodes. Figure~\ref{fig:psa} illustrates this asynchronous approach, which eliminates worker idle time and potentially increases iteration throughput. However, this methodology may introduce parameter inconsistencies due to temporal delays between gradient computation and parameter application, requiring careful consideration of staleness tolerance in the optimization algorithm.

\subsection{Implementation Details and Hyperparameters}
To ensure reproducibility and fair comparison, consistent hyperparameters were maintained across distributed strategies where applicable.

\begin{itemize}
    \item \textbf{Optimization:} All models were trained using the SGD optimizer with an initial learning rate of 0.01 and momentum of 0.9.
    \item \textbf{Batch Size:} A global batch size of 128 was utilized for CIFAR-10 experiments, and 64 for MNIST.
    \item \textbf{Precision:} Experiments were conducted in mixed-precision (FP16) to mirror modern production environments and optimize memory usage.
    \item \textbf{Software Stack:} All experiments were implemented using PyTorch 2.4.0 and CUDA 12.1.
\end{itemize}

\section{Experimental Setup}\label{sec4}

This study employs a dual-cluster experimental framework to comprehensively evaluate distributed deep learning strategies under varying hardware configurations and computational constraints. The experimental design encompasses two distinct computational environments, each optimized for specific distributed training paradigms and model architectures. Table~\ref{tab:experimental_setup} provides a comprehensive overview of the experimental configurations, hardware specifications, and evaluated models across both experimental environments.

\begin{table}[htbp]
\caption{Experimental Setup Details}\label{tab:experimental_setup}%
\begin{tabular}{@{}lll@{}}
\toprule
 & \textbf{Experiment 1}  & \textbf{Experiment 2} \\
\midrule
\textbf{Platform} & University Hopper Cluster & Lightning.AI Cluster \\
\midrule
\textbf{Strategies} & \makecell[l]{DDP \\ FSDP} & \makecell[l]{Synchronous-PS \\ Asynchronous-PS} \\
\midrule
\textbf{Hardware} & \makecell[l]{1x, 2x, 4x A100.80GB GPUs} & \makecell[l]{1x, 2x, 4x A10G.24GB GPUs} \\
\midrule
\textbf{RAM} & 500 GB & 200 GB \\
\midrule
\textbf{CPUs} & 64 & 48 \\
\midrule
\textbf{Models} & \makecell[l]{VGG16\\EfficientNet\_v2\\ConvNext\_Large} & \makecell[l]{ResNet50\\DenseNet121\\VGG16} \\
\midrule
\textbf{Dataset} & CIFAR-10 & MNIST \\
\botrule
\end{tabular}
\end{table}

\begin{figure}[ht]
\centering
\includegraphics[width=0.9\textwidth]{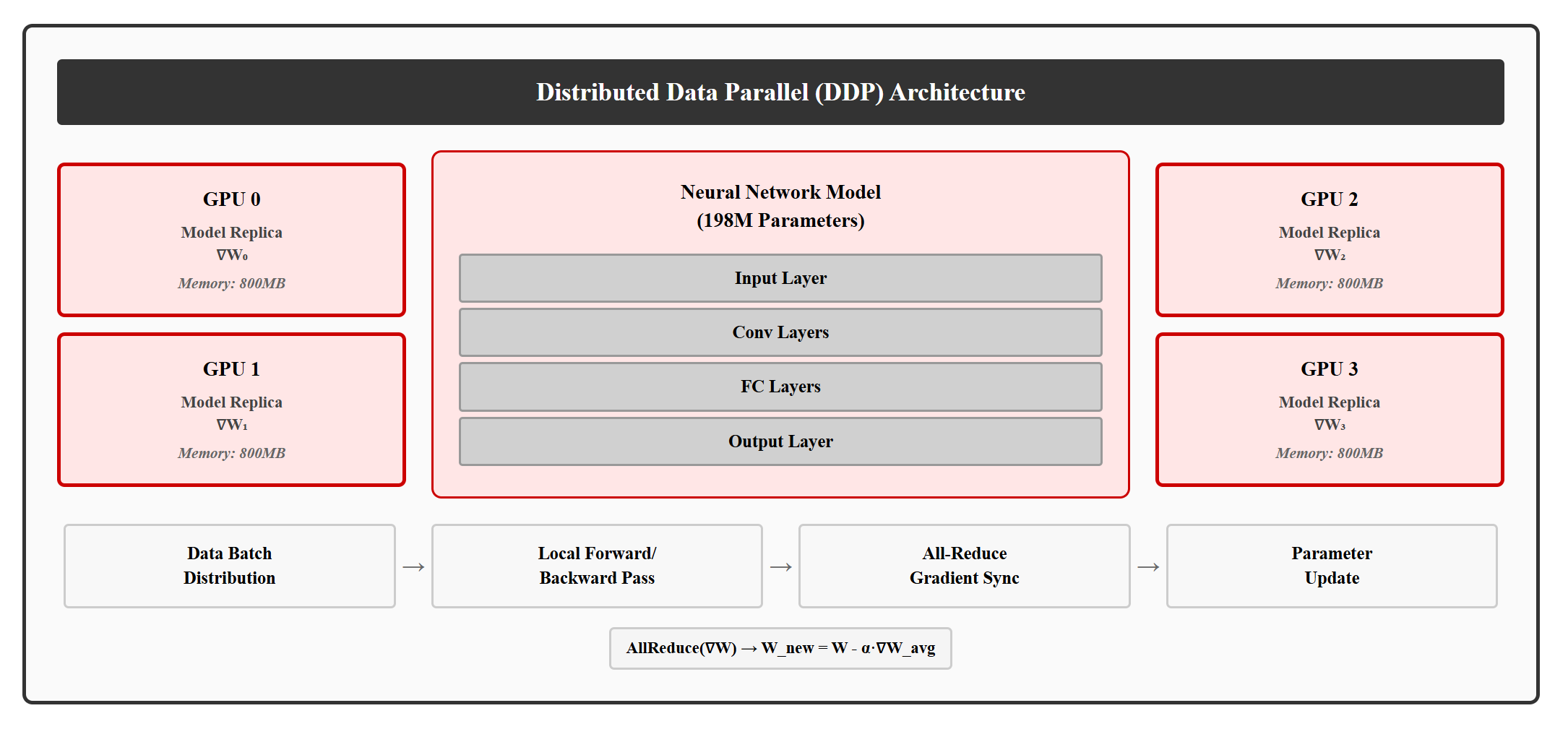}
\caption{DDP architecture showing complete model replication across GPUs with an all-reduce gradient synchronization pattern. Each GPU maintains a full model replica, requiring approximately 800MB of memory per device (for VGG16).}\label{fig:ddp_arch}
\end{figure}

\begin{figure}[ht]
\centering
\includegraphics[width=0.9\textwidth]{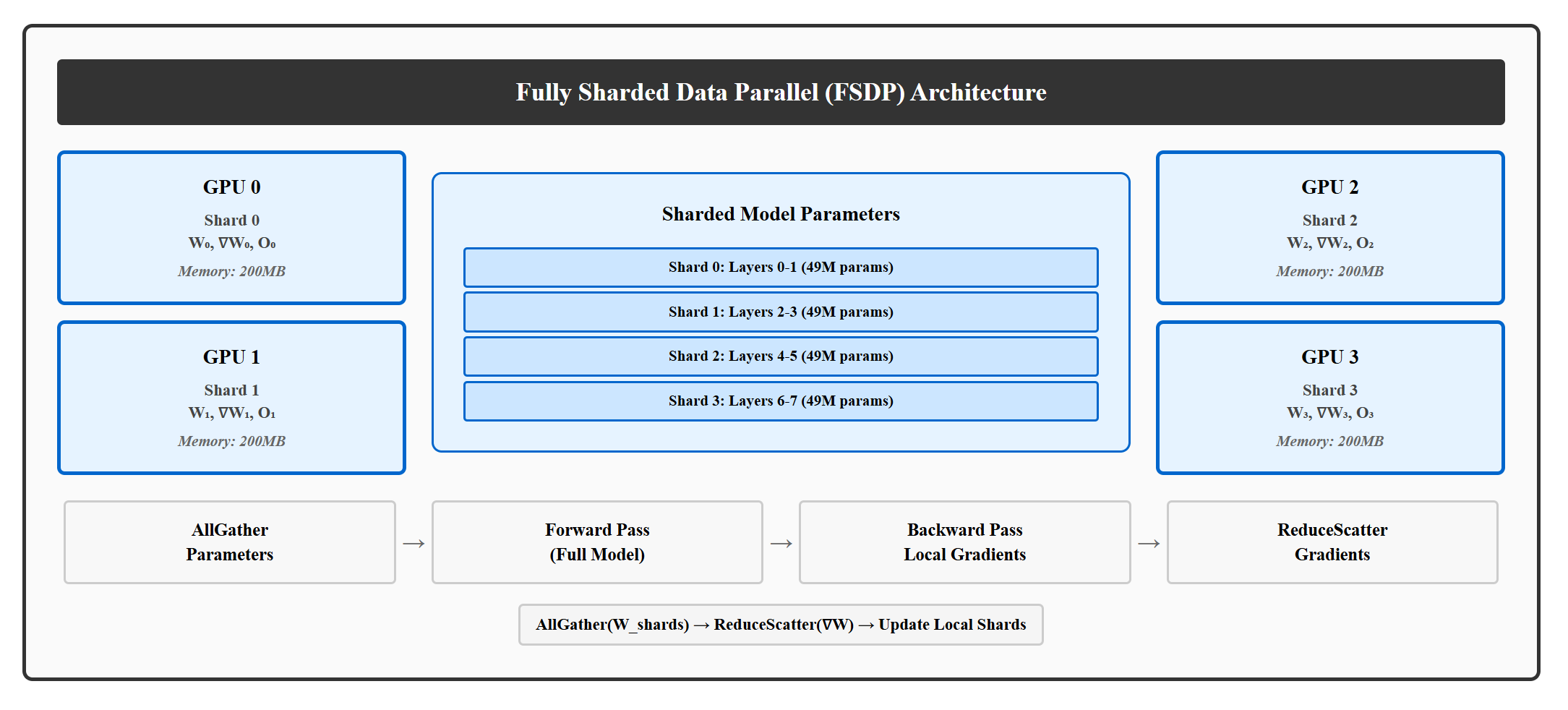}
\caption{FSDP architecture illustrating parameter sharding across GPUs. The forward pass uses all-gather while the backward pass uses reduce-scatter operations. Each shard contains approximately 49 million parameters (for VGG16), with a reduced memory footprint of around 200MB per GPU.}\label{fig:fsdp_arch}
\end{figure}

\begin{figure}[ht]
\centering
\includegraphics[width=0.9\textwidth]{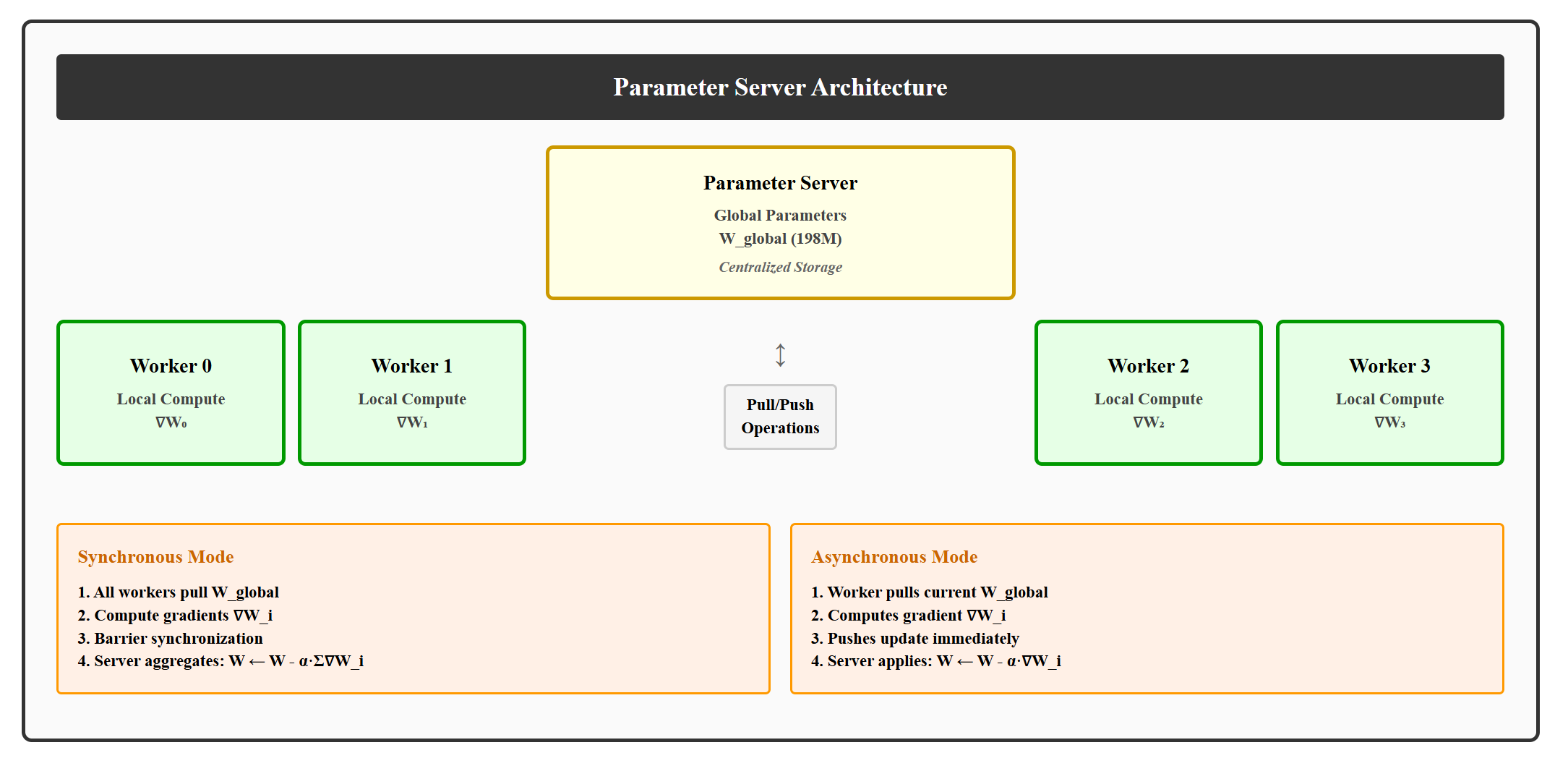}
\caption{Parameter Server architecture featuring centralized parameter storage and distributed worker nodes. Workers use pull/push communication to exchange parameters with the server. The diagram highlights both synchronous and asynchronous update modes.}\label{fig:ps_arch}
\end{figure}

The experimental framework systematically evaluates three distinct distributed training paradigms through their respective architectural implementations. Figure~\ref{fig:ddp_arch} illustrates the Distributed Data Parallel architecture, demonstrating complete model replication across GPUs with all-reduce gradient synchronization patterns. Each GPU maintains a full model replica, requiring approximately 800MB of memory per device (for VGG16). Figure~\ref{fig:fsdp_arch} depicts the Fully Sharded Data Parallel architecture, showcasing parameter sharding across GPUs where the forward pass employs all-gather operations while the backward pass utilizes reduce-scatter operations. Each shard contains approximately 49 million parameters (for VGG16), achieving a reduced memory footprint of around 200MB per GPU. Figure~\ref{fig:ps_arch} presents the Parameter Server architecture featuring centralized parameter storage and distributed worker nodes, where workers utilize pull/push communication mechanisms to exchange parameters with the server, highlighting both synchronous and asynchronous update modes.

\subsection{Environment 1: University Hopper Cluster}\label{subsec4_1}

The first experimental environment utilizes the University Hopper Cluster, equipped with high-performance NVIDIA A100 GPUs featuring 80GB of High Bandwidth Memory (HBM) per device. This environment enables evaluation of memory-intensive models under 1 GPU, 2-GPU and 4-GPU configurations, providing 500GB of system RAM and 64 CPU cores for computational support.

The experimental protocol implements both DDP and FSDP strategies as model wrappers around three architectures: VGG16, EfficientNet-v2, and ConvNext-Large. In the DDP implementation, each model is completely replicated across participating GPUs, with PyTorch's DistributedDataParallel module managing gradient synchronization through NCCL all-reduce operations. The FSDP implementation employs PyTorch's native FSDP wrapper, which automatically shards model parameters, gradients, and optimizer states across GPUs while maintaining computational correctness through dynamic parameter gathering and scattering operations.

Training proceeds for 10 epochs using the CIFAR-10 dataset, with systematic measurement of total training time, per-GPU utilization metrics, and memory consumption patterns. The experimental design enables direct comparison of synchronous distributed strategies under identical computational and data conditions, isolating the performance impact of different parallelization approaches.

\subsection{Environment 2: Lightning.AI Cluster}\label{subsec4_2}

The second experimental environment leverages the Lightning.AI computational platform \cite{b30}, configured with NVIDIA A10G GPUs featuring 24GB of memory per device. This environment provides 200GB of system RAM and 48 CPU cores, creating a resource-constrained scenario suitable for evaluating Parameter Server architectures.

The Parameter Server implementation establishes a distributed architecture where model parameters are centrally maintained on dedicated server nodes while worker nodes execute gradient computation using allocated data partitions. The synchronous Parameter Server configuration ensures all workers complete gradient computation before parameter updates are applied, maintaining model consistency at the expense of computational efficiency. The asynchronous variant permits independent worker operation, allowing continuous parameter updates without synchronization barriers.

Three model architectures (ResNet50, DenseNet121, and VGG16) are evaluated under both Parameter Server configurations using the MNIST dataset. The experimental protocol systematically varies the number of participating GPUs (1, 2, and 4 configurations) to assess scalability characteristics. Training metrics including total training time, model accuracy, GPU utilization, and memory usage are recorded across all configurations.

Additional optimization experiments are conducted for the asynchronous Parameter Server strategy, incorporating adjustments to communication frequency and gradient update methodologies to enhance performance while maintaining convergence stability. These modifications aim to mitigate the adverse effects of parameter staleness inherent in asynchronous distributed training while preserving the computational efficiency advantages of reduced synchronization overhead.

\section{Results and Analysis}\label{sec5}

\subsection{Experiment 1: DDP versus FSDP Performance Analysis}

The first experimental evaluation examines the comparative performance of Distributed Data Parallel (DDP) and Fully Sharded Data Parallel (FSDP) strategies on the University Hopper Cluster. This analysis encompasses three neural network architectures (VGG16, EfficientNet-v2, and ConvNeXt-Large) trained on the CIFAR-10 dataset across three GPU configurations (1, 2, and 4 GPUs) to assess scalability characteristics and resource utilization patterns.

\begin{table}[htbp]
\caption{Training Time (in seconds) for DDP and FSDP Strategies}\label{tab:training_time_restructured}
\begin{tabular}{@{}lllll@{}}
\toprule
\textbf{Strategy} & \textbf{Model} & \textbf{1 GPU} & \textbf{2 GPUs} & \textbf{4 GPUs} \\
\midrule
\multirow{3}{*}{DDP} 
& VGG16 & 487.55 & 230.74 & 113.81 \\
& EfficientNet\_v2 & 1786.19 & 836.28 & 365.63 \\
& ConvNeXt\_Large & 1678.77 & 794.65 & 485.89 \\
\midrule
\multirow{3}{*}{FSDP} 
& VGG16 & 1059.67 & 490.32 & 279.53 \\
& EfficientNet\_v2 & 10292.84 & 4882.40 & 2110.95 \\
& ConvNeXt\_Large & 5891.63 & 2786.49 & 1490.43 \\
\bottomrule
\end{tabular}
\end{table}

\subsubsection{Training Time Performance}

Table~\ref{tab:training_time_restructured} presents comprehensive training time measurements across all experimental configurations. The results demonstrate a consistent pattern wherein FSDP requires substantially longer training durations compared to DDP across all model architectures and GPU configurations. Specifically, FSDP exhibits training times that are approximately two to three times greater than corresponding DDP implementations across all three GPU configurations (1, 2, and 4 GPUs).

For the VGG16 architecture, the single-GPU baseline shows DDP achieving training completion in 487.55 seconds, while FSDP requires 1059.67 seconds, representing a 2.17× increase in training duration. When scaling to 2 GPUs, DDP reduces training time to 230.74 seconds while FSDP achieves 490.32 seconds, maintaining the approximately 2× performance differential. The 4-GPU configuration further demonstrates this consistent pattern, with DDP completion times of 113.81 seconds versus 279.53 seconds for FSDP.

EfficientNet-v2 exhibits more pronounced differences, with FSDP showing training time increases of 5.76× (single GPU), 5.84× (2 GPUs), and 5.77× (4 GPUs) compared to corresponding DDP implementations. ConvNeXt-Large demonstrates similar patterns with FSDP requiring 3.51×, 3.51×, and 3.07× longer training times for 1, 2, and 4 GPU configurations respectively.

The scaling efficiency analysis reveals consistent performance improvements when increasing GPU count across both strategies. Transitioning from single GPU to 2-GPU configurations reduces training time by approximately 50\% for both DDP and FSDP across all model architectures. Further scaling from 2 to 4 GPUs continues this trend, with DDP maintaining near-linear scaling properties and FSDP demonstrating similar scaling efficiency despite higher absolute training times.

\begin{figure}[htbp]
\centering
\includegraphics[width=\textwidth]{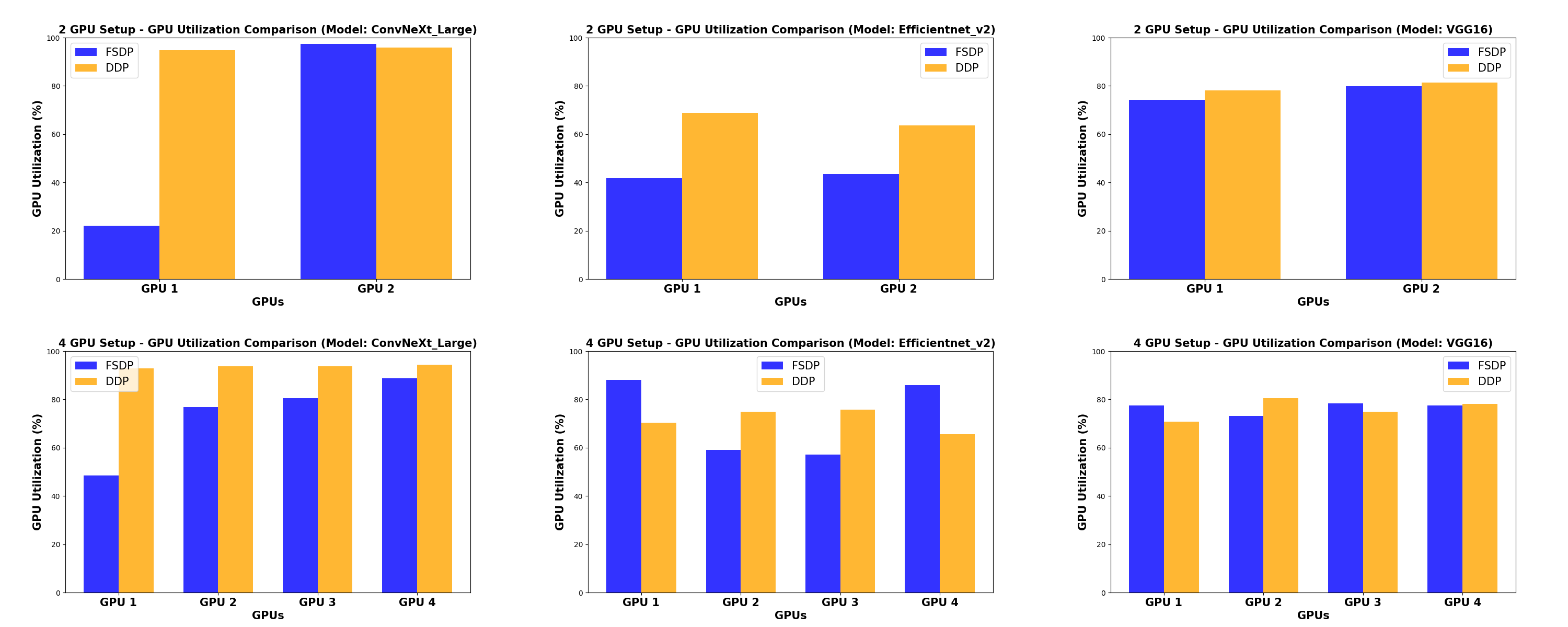}
\caption{GPU Utilization Comparison between DDP and FSDP Strategies}\label{fig:ddp_fsdp_gpu}
\end{figure}

\begin{figure}[htbp]
\centering
\includegraphics[width=\textwidth]{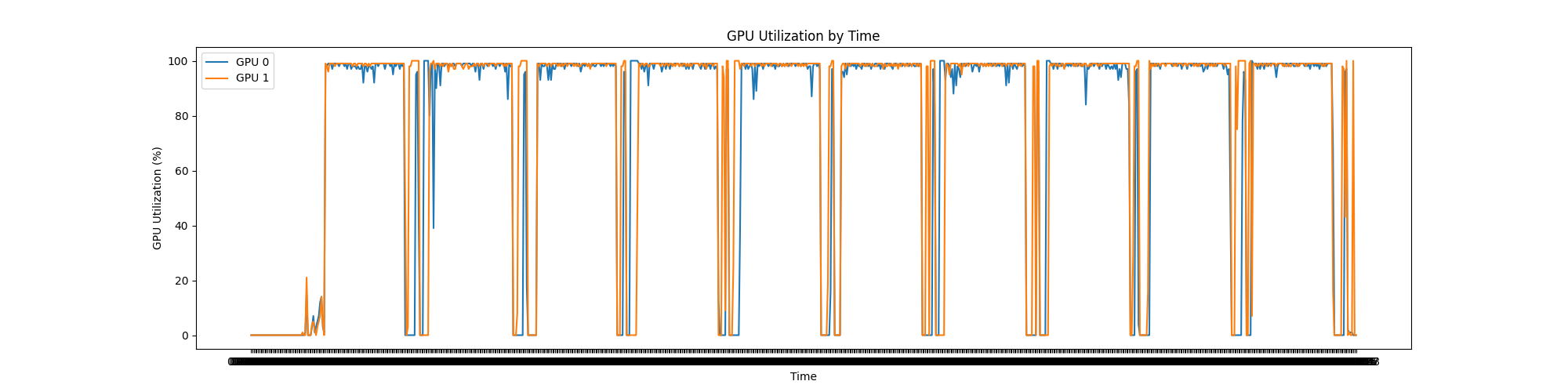}
\caption{GPU Utilization for DDP using 2 GPUs}
\label{fig:gpu_ddp2}
\end{figure}

\begin{figure}[htbp]
\centering
\includegraphics[width=\textwidth]{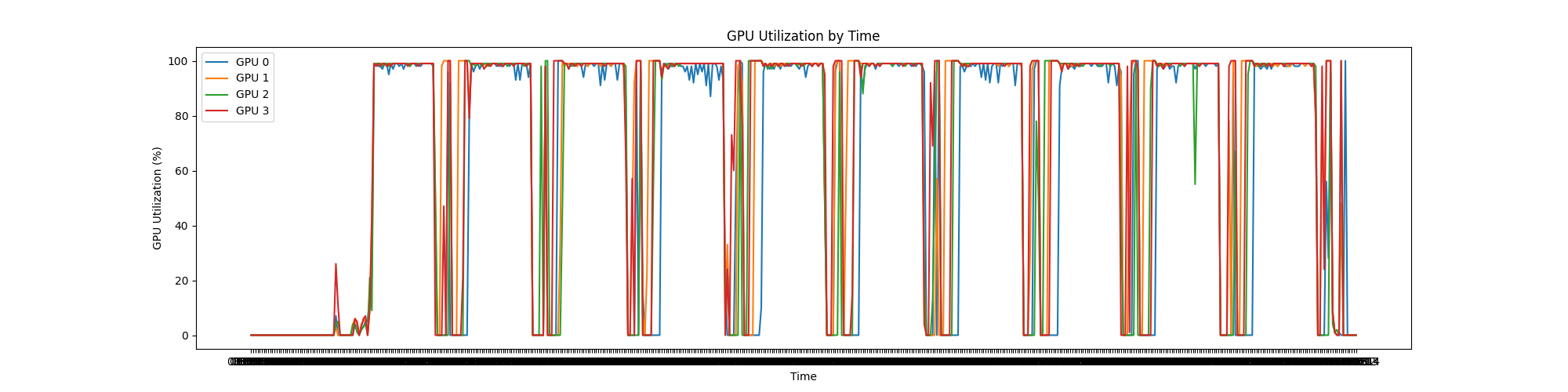}
\caption{GPU Utilization for DDP using 4 GPUs}
\label{fig:gpu_ddp4}
\end{figure}

\begin{figure}[htbp]
\centering
\includegraphics[width=\textwidth]{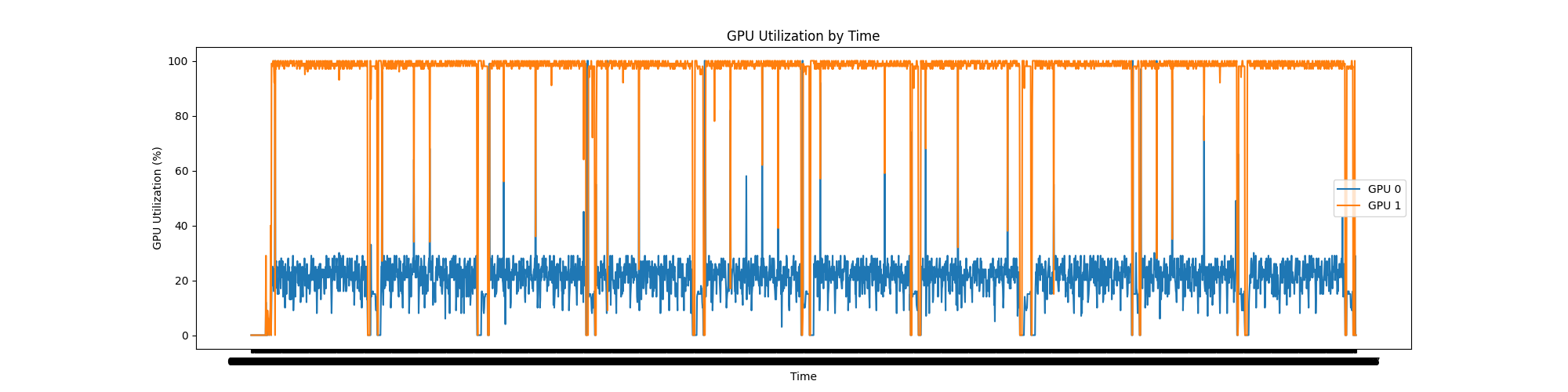}
\caption{GPU Utilization for FSDP using 2 GPUs}
\label{fig:gpu_fsdp2}
\end{figure}

\begin{figure}[htbp]
\centering
\includegraphics[width=\textwidth]{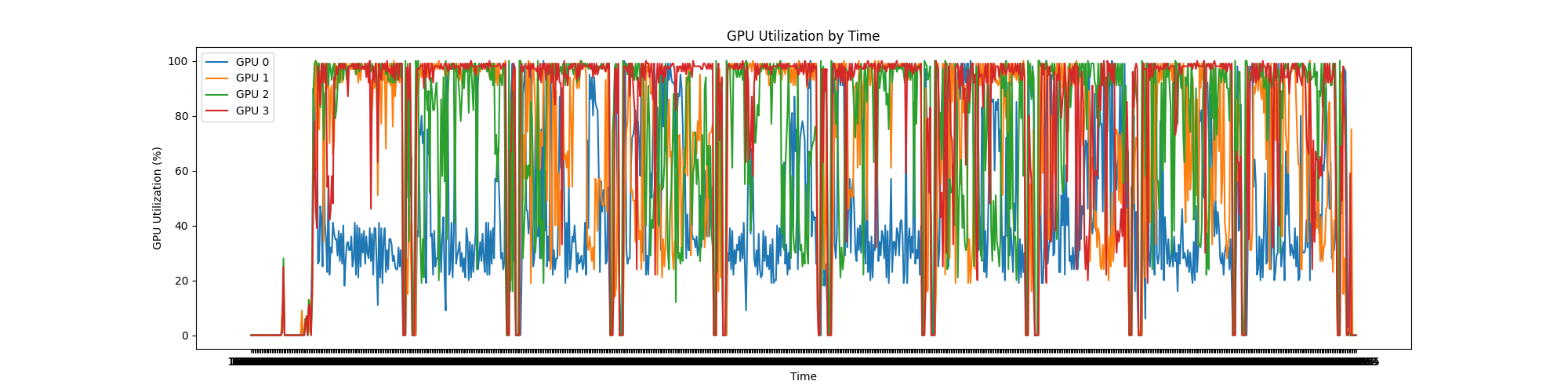}
\caption{GPU Utilization for FSDP using 4 GPUs}
\label{fig:gpu_fsdp4}
\end{figure}

\subsubsection{GPU Utilization Analysis}

Figures~\ref{fig:gpu_ddp2}, \ref{fig:gpu_ddp4}, \ref{fig:gpu_fsdp2}, and \ref{fig:gpu_fsdp4} present detailed GPU utilization patterns for both strategies across 2-GPU and 4-GPU configurations during ConvNeXt-Large training. The comparative analysis illustrated in Figure~\ref{fig:ddp_fsdp_gpu} demonstrates that GPU utilization metrics remain comparable between DDP and FSDP strategies across all evaluated configurations, with both approaches achieving similar computational efficiency levels.

The GPU utilization profiles indicate consistent processing loads across participating devices for both 2-GPU and 4-GPU configurations, suggesting that both strategies effectively distribute computational workloads without significant load imbalance. The absence of substantial utilization differences between DDP and FSDP implementations indicates that the observed training time disparities are attributable to communication overhead and synchronization patterns rather than computational inefficiencies.

\begin{figure}[htbp]
\centering
\includegraphics[width=\textwidth]{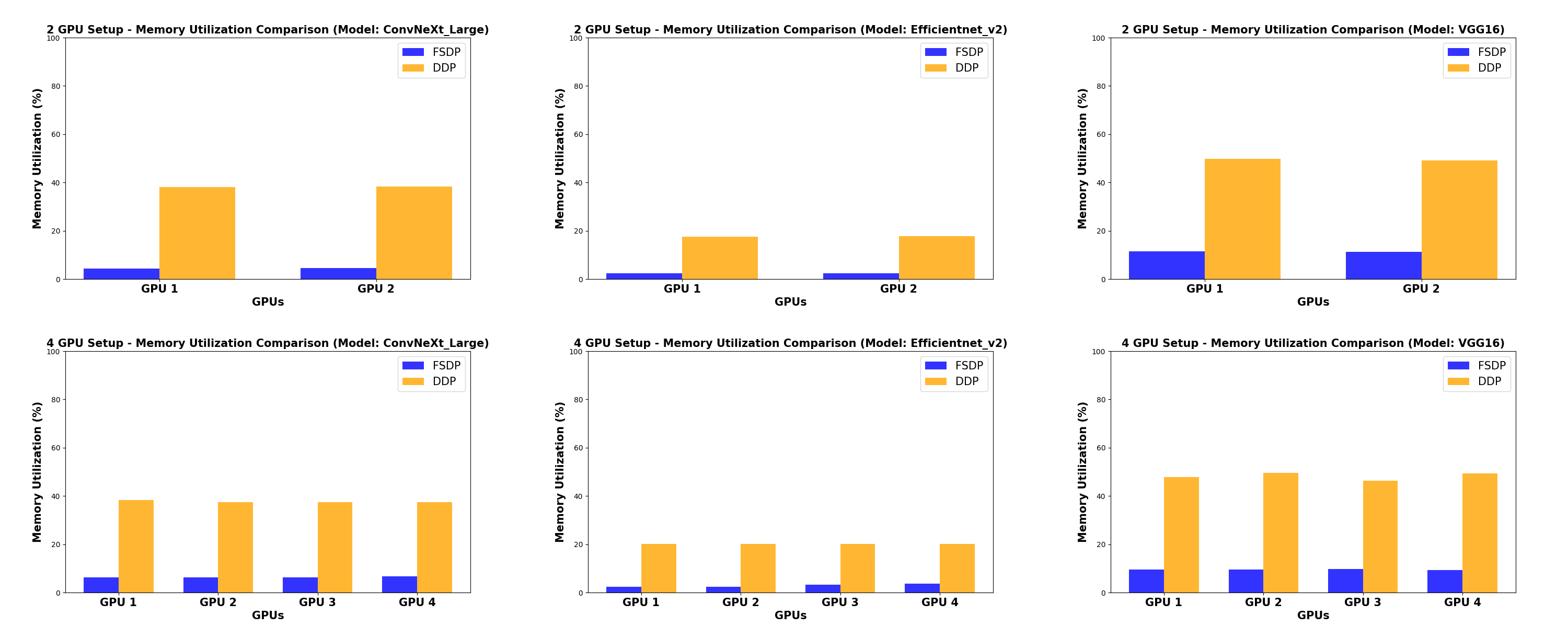}
\caption{Memory Utilization Comparison between DDP and FSDP Strategies}\label{fig:ddp_fsdp_mem}
\end{figure}

\begin{figure}[htbp]
\centering
\includegraphics[width=\textwidth]{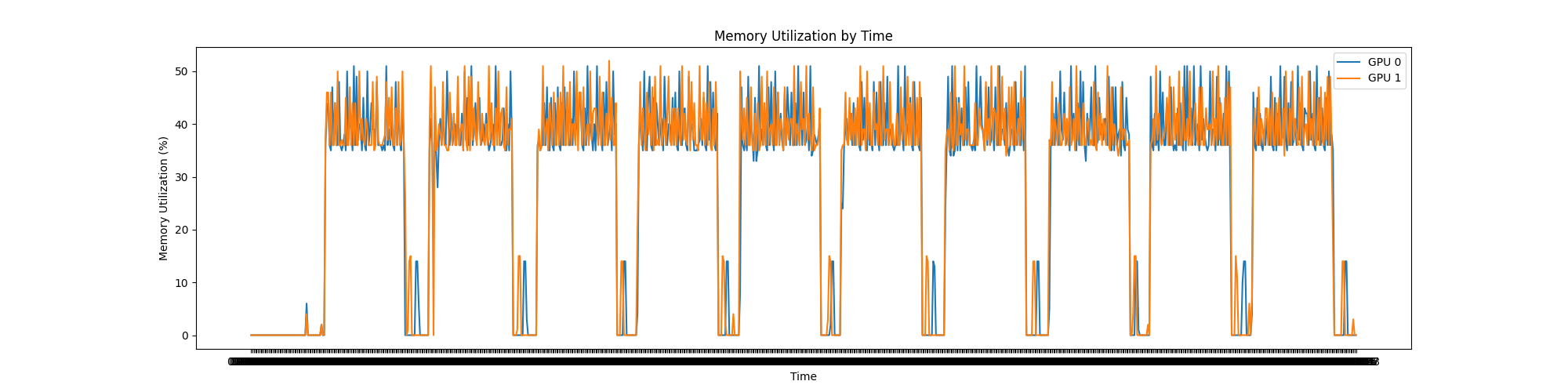}
\caption{Memory Utilization for DDP using 2 GPUs}
\label{fig:memory_ddp2}
\end{figure}

\begin{figure}[htbp]
\centering
\includegraphics[width=\textwidth]{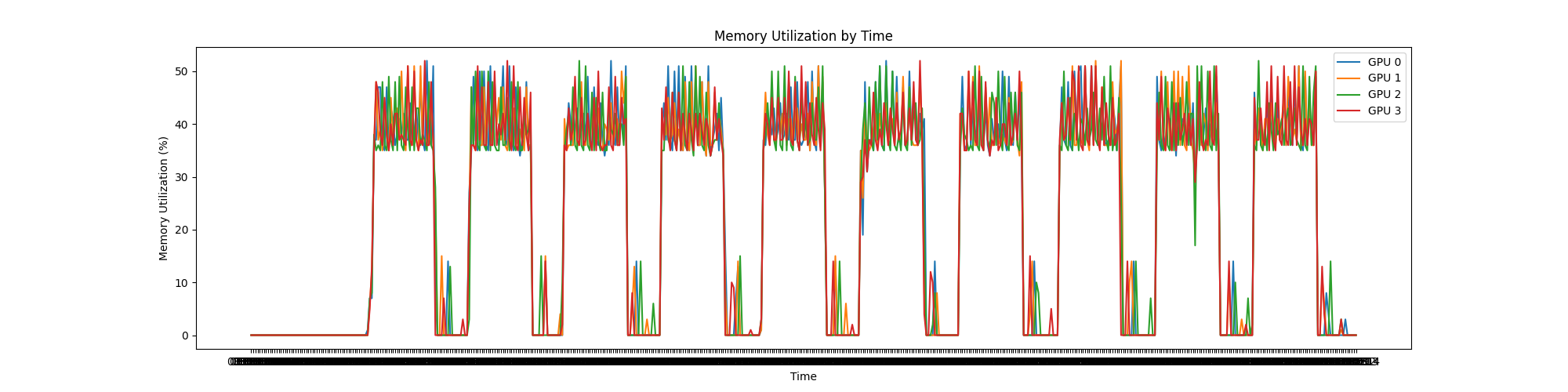}
\caption{Memory Utilization for DDP using 4 GPUs}
\label{fig:memory_ddp4}
\end{figure}

\begin{figure}[htbp]
\centering
\includegraphics[width=\textwidth]{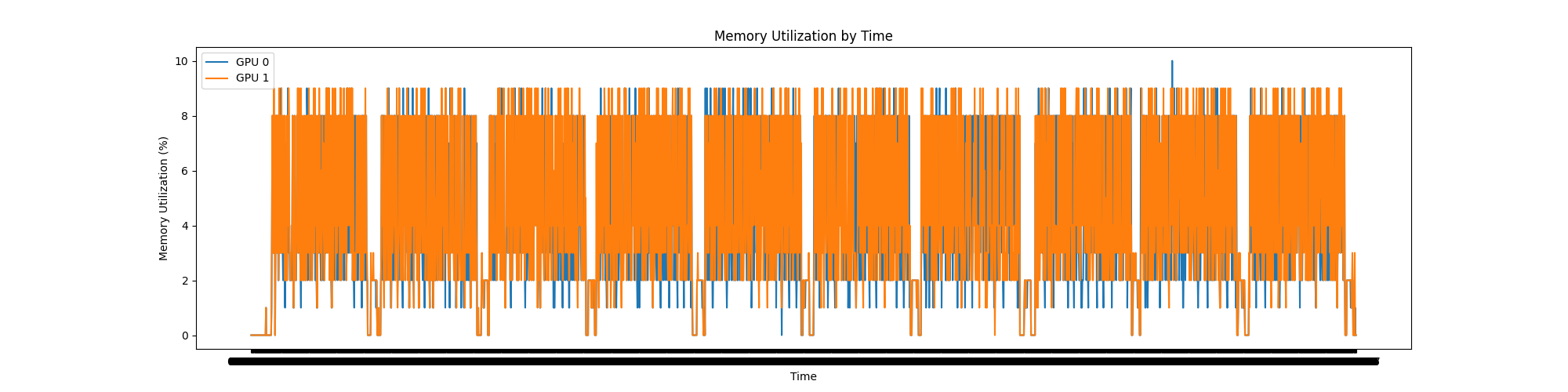}
\caption{Memory Utilization for FSDP using 2 GPUs}
\label{fig:memory_fsdp2}
\end{figure}

\begin{figure}[htbp]
\centering
\includegraphics[width=\textwidth]{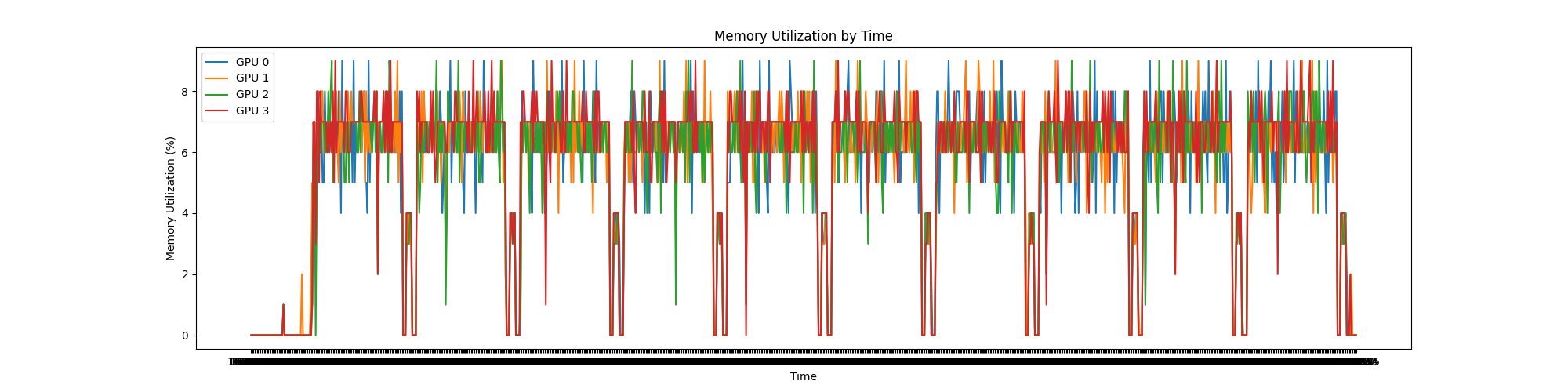}
\caption{Memory Utilization for FSDP using 4 GPUs}
\label{fig:memory_fsdp4}
\end{figure}

\subsubsection{Memory Utilization Characteristics}

The memory utilization analysis reveals the most significant performance differentiation between DDP and FSDP strategies across all GPU configurations. Figures~\ref{fig:memory_ddp2}, \ref{fig:memory_ddp4}, \ref{fig:memory_fsdp2}, and \ref{fig:memory_fsdp4} demonstrate substantial differences in per-GPU memory consumption patterns during ConvNeXt-Large training across 2-GPU and 4-GPU configurations.

Figure~\ref{fig:ddp_fsdp_mem} provides a comprehensive comparison of memory utilization between the two strategies, clearly illustrating that FSDP achieves memory consumption reductions of four to six times compared to DDP implementations across all evaluated configurations. This substantial memory efficiency improvement stems from FSDP's parameter sharding mechanism, which distributes model parameters across participating GPUs rather than maintaining complete replicas on each device.

The memory efficiency gains become particularly pronounced for larger model architectures. While DDP requires storing full model replicas, FSDP significantly reduces the per-device footprint, demonstrating the potential to scale to much larger models that would otherwise exceed single-GPU capacity.

\subsubsection{Performance Trade-off Analysis}

The experimental results reveal a fundamental trade-off between training time efficiency and memory resource utilization when comparing DDP and FSDP strategies across all three GPU configurations. While DDP demonstrates superior training time performance through reduced communication overhead and simplified synchronization mechanisms, FSDP provides substantial memory efficiency advantages that enable training of larger models under memory-constrained scenarios.

The consistent 2-3× training time increase observed with FSDP across 1, 2, and 4 GPU configurations can be attributed to the additional communication overhead required for parameter gathering and scattering operations during forward and backward propagation phases. However, the 4-6× reduction in memory consumption achieved by FSDP represents a significant advantage for scenarios where memory constraints limit model size or batch size selections.

These findings indicate that the selection between DDP and FSDP strategies should be informed by the specific constraints and objectives of the training scenario. DDP remains optimal for environments with sufficient memory resources where training time minimization is prioritized, while FSDP becomes advantageous when memory efficiency is critical for enabling larger model training or accommodating hardware limitations across any GPU configuration scale.

\subsection{Experiment 2: Synchronous versus Asynchronous Parameter Server Analysis}

The second experimental evaluation investigates the performance characteristics of synchronous and asynchronous Parameter Server (PS) strategies on the Lightning.AI cluster. This analysis examines three neural network architectures (DenseNet121, ResNet50, and VGG16) trained on the MNIST dataset across three GPU configurations (1, 2, and 4 GPUs) to assess the trade-offs between training speed and model accuracy under different synchronization paradigms.

A key goal of this experiment is \textit{system-level characterization} rather than maximizing the absolute accuracy of MNIST classifiers. To isolate the effects of synchronization, we intentionally used a fixed training recipe across PS modes and GPU scales (same optimizer family, learning rate, momentum, epoch budget, data pipeline, and evaluation procedure). We did not introduce scale-specific tuning such as learning rate scaling, warmup, gradient clipping, per-worker batch resizing, or staleness-aware optimizer adjustments. As a result, the absolute accuracy values should be interpreted as a proxy for \textit{optimization stability under a fixed configuration}, not as the best achievable MNIST performance for these architectures.

\begin{table}[htbp]
\caption{Accuracy (\%) for Parameter Server Strategies Across GPU Configurations}\label{tab:accuracy_comparison_restructured}
\begin{tabular}{@{}lllll@{}}
\toprule
\textbf{Strategy} & \textbf{Model} & \textbf{1 GPU} & \textbf{2 GPUs} & \textbf{4 GPUs} \\
\midrule
\multirow{3}{*}{Synchronous}
& DenseNet   & 51.6 & 49.2 & 46.3 \\
& ResNet50   & 52.2 & 47.6 & 40.2 \\
& VGG16      & 98.4 & 96.9 & 47.6 \\
\midrule
\multirow{3}{*}{Asynchronous}
& DenseNet   & 45.3 & 38.3 & 28.4 \\
& ResNet50   & 43.9 & 38.8 & 34.7 \\
& VGG16      & 83.2 & 78.2 & 42.4 \\
\bottomrule
\end{tabular}
\end{table}

\begin{table}[htbp]
\caption{Training Time (in seconds) for Parameter Server Strategies Across GPU Configurations}\label{tab:training_time_ps_restructured}
\begin{tabular}{@{}lllll@{}}
\toprule
\textbf{Strategy} & \textbf{Model} & \textbf{1 GPU} & \textbf{2 GPUs} & \textbf{4 GPUs} \\
\midrule
\multirow{3}{*}{Synchronous}
& DenseNet   & 509.88 & 231.04 & 137.34 \\
& ResNet50   & 468.31 & 246.54 & 129.24 \\
& VGG16      & 891.52 & 545.61 & 247.66 \\
\midrule
\multirow{3}{*}{Asynchronous}
& DenseNet   & 506.42 & 219.57 & 117.43 \\
& ResNet50   & 464.91 & 220.97 & 109.02 \\
& VGG16      & 680.22 & 351.62 & 177.86 \\
\bottomrule
\end{tabular}
\end{table}

\subsubsection{Training Time Performance}

Table~\ref{tab:training_time_ps_restructured} presents comprehensive training time measurements for both synchronous and asynchronous Parameter Server configurations across all experimental setups. The results demonstrate that asynchronous Parameter Server configurations consistently achieve superior training time performance compared to their synchronous counterparts across all model architectures and GPU configurations.

For the DenseNet architecture, asynchronous Parameter Server achieves training completion in 506.42 seconds on a single GPU compared to 509.88 seconds for the synchronous configuration, representing a modest 0.68\% improvement. However, the performance advantage becomes more pronounced with increased GPU scaling. In the 4-GPU configuration, asynchronous Parameter Server completes training in 117.43 seconds compared to 137.34 seconds for synchronous implementation, demonstrating a 14.5\% reduction in training time.

The VGG16 architecture exhibits the most significant performance differentials between synchronous and asynchronous configurations. In the 4-GPU setup, asynchronous Parameter Server achieves completion in 177.86 seconds compared to 247.66 seconds for synchronous implementation, representing a 28\% improvement in training efficiency. ResNet50 demonstrates similar patterns, with asynchronous configurations achieving 15.6\% faster training times in the 4-GPU configuration (109.02 seconds versus 129.24 seconds).

The scaling analysis reveals that both synchronous and asynchronous Parameter Server strategies maintain effective scaling properties when increasing GPU count from 1 to 4 devices. However, asynchronous configurations demonstrate superior scaling efficiency, achieving greater training time reductions as the number of participating GPUs increases. This behavior is expected because asynchronous execution removes the strict iteration barrier, reducing idle time caused by worker speed variation and communication delays.

\subsubsection{Model Accuracy Performance}

Table~\ref{tab:accuracy_comparison_restructured} presents model accuracy measurements across all Parameter Server configurations, revealing a critical trade-off between training speed and final model performance. The results demonstrate that while asynchronous Parameter Server configurations achieve faster training times, they consistently produce lower model accuracy across all architectures and GPU configurations.

For the VGG16 architecture, synchronous Parameter Server achieves 98.4\% accuracy in single-GPU configuration, while asynchronous implementation attains 83.2\% accuracy, representing a 15.2 percentage point reduction. This accuracy differential persists across scaling scenarios, with 4-GPU configurations showing synchronous accuracy of 47.6\% versus 42.4\% for asynchronous implementation.

DenseNet and ResNet50 architectures exhibit similar patterns, with asynchronous configurations consistently underperforming synchronous implementations. In the 4-GPU configuration, DenseNet accuracy decreases from 46.3\% (synchronous) to 28.4\% (asynchronous), while ResNet50 accuracy drops from 40.2\% to 34.7\%. These accuracy reductions are consistent with \textit{gradient and parameter staleness} in asynchronous PS: workers compute gradients using parameter versions that may already have been updated multiple times by other workers, so the applied updates are less aligned with the current optimization state.

Beyond staleness, the experiment shows an accuracy decline as GPU count increases for both synchronous and asynchronous PS across all three models. Under our fixed training recipe, this trend is explained by the following interacting factors.

\textbf{(1) Effective global batch and step dynamics under scaling.} When scaling from 1 to 2 and 4 GPUs, the effective amount of data processed per unit time increases substantially. If the per-worker batch size is held constant, the implicit global batch size grows with the number of workers. If instead the global batch is held fixed, then each worker sees fewer samples per step, changing gradient noise statistics and update cadence. In both cases, keeping the same learning rate schedule and epoch budget can lead to under-training, reduced generalization, or unstable convergence. In practical distributed training, these effects are commonly mitigated with learning rate scaling rules, warmup, and schedule adjustments, which were intentionally not applied here to preserve experimental comparability.

\textbf{(2) Communication delay and asynchronous version skew.} As the number of workers increases, communication contention and queueing at the parameter server can increase. In synchronous PS this manifests as longer waits at iteration boundaries and potential straggler amplification. In asynchronous PS it manifests as larger parameter version skew, meaning workers are more likely to compute gradients on older parameters. This increases gradient inconsistency and can degrade convergence even if wall-clock training is faster.

\textbf{(3) Short training horizon by design.} The PS experiment is designed to stress synchronization behavior under a limited, consistent epoch budget. Under such a constrained horizon, optimization noise from staleness and scaling effects has less opportunity to average out, which can lead to depressed final accuracies compared to longer, tuned training.

\textbf{(4) Model-specific sensitivity.} The sharp drop observed for VGG16 at 4 GPUs (for both synchronous and asynchronous PS) indicates that VGG16 is particularly sensitive to the fixed hyperparameter and synchronization regime in this setup. Importantly, this observation is reported as an empirical \textit{failure mode of naive scaling in PS} rather than a claim about the intrinsic performance of VGG16 on MNIST. In production PS deployments, stabilizing measures such as tuned learning rate schedules, gradient clipping, server-side momentum, bounded staleness, or partial synchronization are typically introduced to avoid this behavior.

Overall, the accuracy results should be interpreted as evidence that PS synchronization choices and scale can materially affect optimization quality under a fixed configuration. The consistent gap between synchronous and asynchronous PS supports the conclusion that removing synchronization barriers improves throughput but increases optimization noise through staleness.

\begin{figure}[htbp]
\centering
\includegraphics[width=\textwidth]{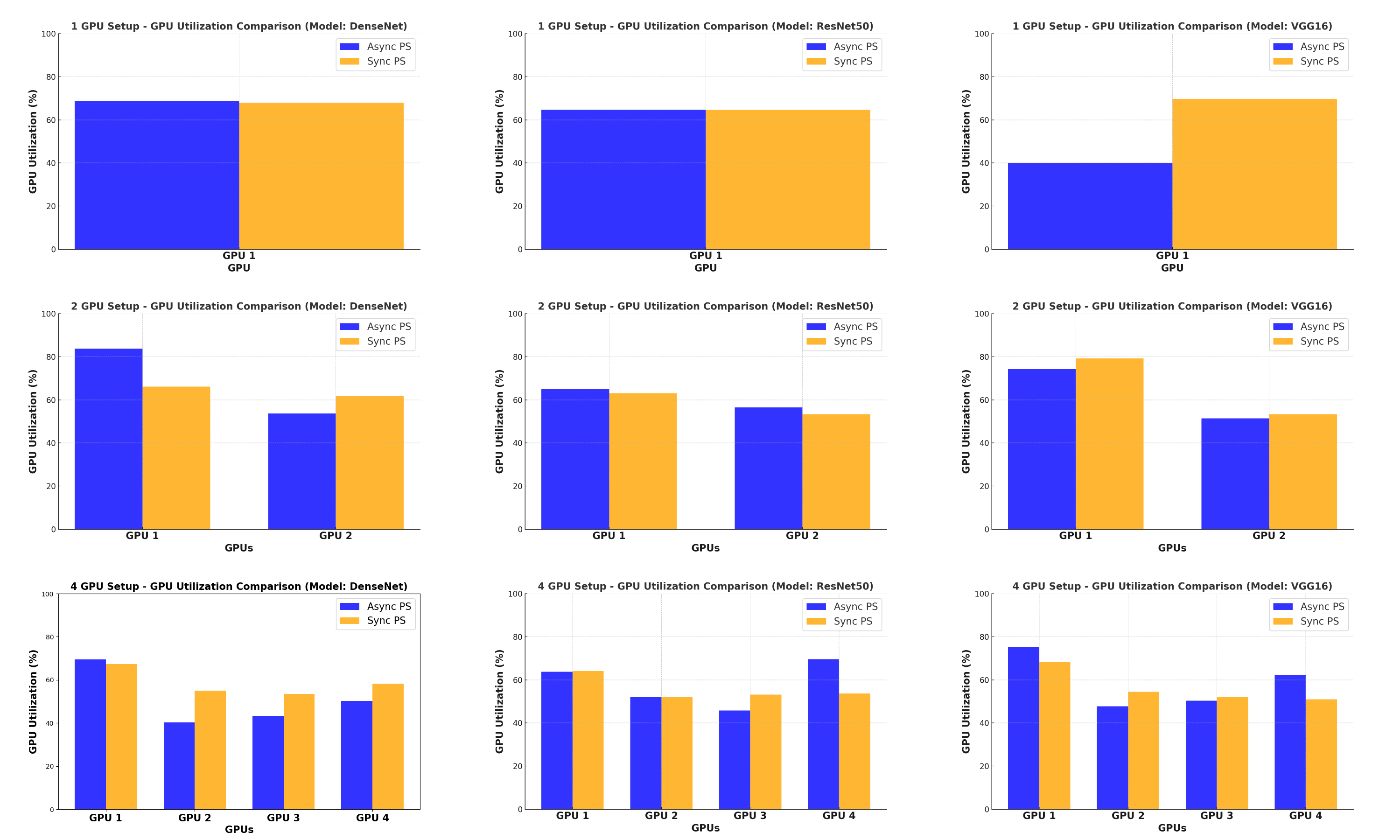}
\caption{GPU Utilization Comparison between different PS Strategies}\label{fig:ps_gpu}
\end{figure}

\begin{figure}[htbp]
\centering
\includegraphics[width=\textwidth]{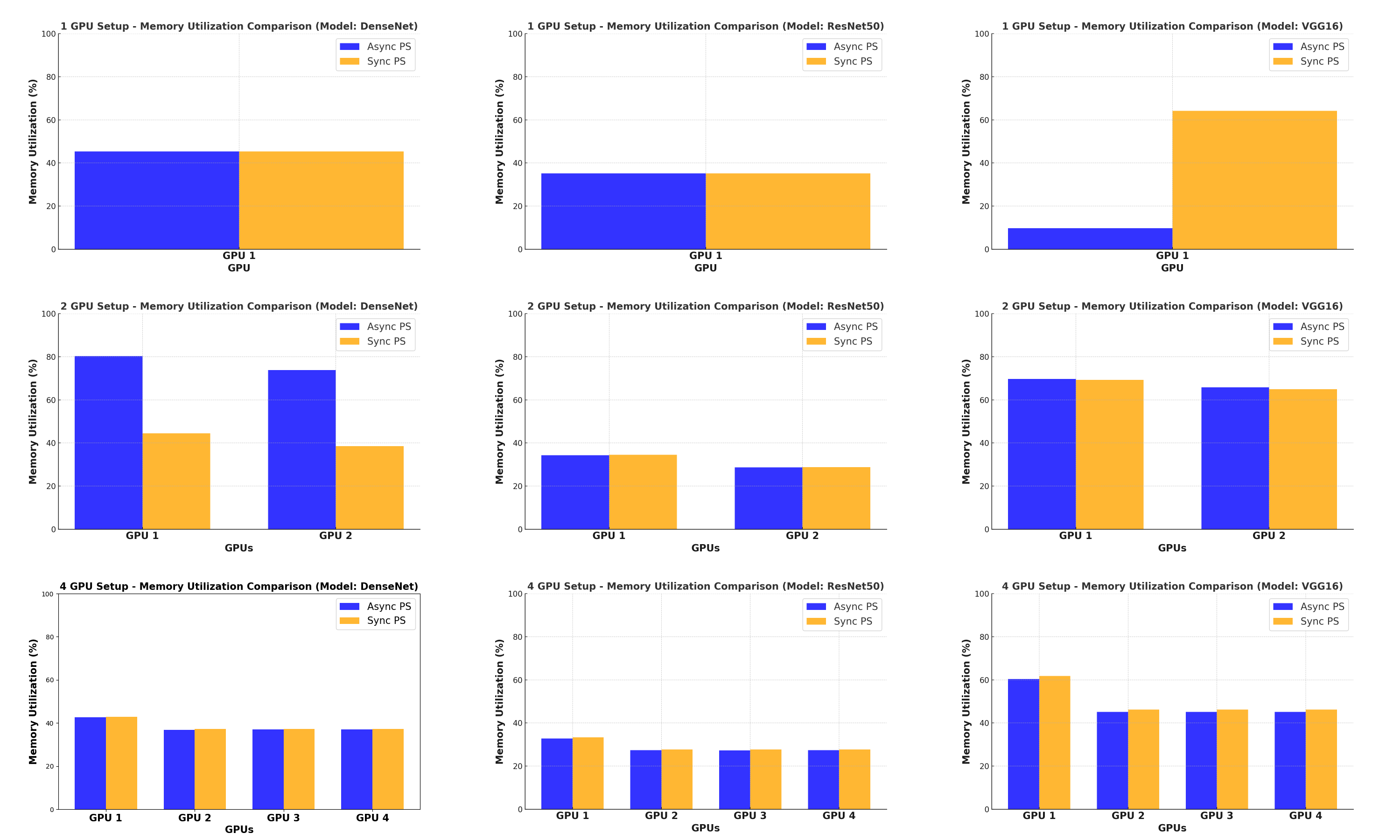}
\caption{Memory Utilization Comparison between different PS Strategies}\label{fig:ps_mem}
\end{figure}

\subsubsection{Resource Utilization Analysis}

Figure~\ref{fig:ps_gpu} illustrates GPU utilization patterns for both synchronous and asynchronous Parameter Server strategies across different configurations. The analysis reveals that synchronous Parameter Server implementations typically achieve higher GPU utilization compared to asynchronous setups. This enhanced utilization efficiency stems from the coordinated computational patterns inherent in synchronous training, where all workers operate in lockstep during gradient computation and parameter update phases.

The synchronous configuration demonstrates more consistent GPU utilization across both single and multi-GPU configurations, indicating effective workload distribution and minimal idle time during training iterations. Conversely, asynchronous configurations exhibit lower overall utilization due to the inherent load imbalance that occurs when workers operate independently without synchronization barriers. Even when utilization is slightly lower, asynchronous PS can still reduce total wall-clock time because it avoids global waiting and continues to make progress while individual workers experience variable compute and communication delays.

Figure~\ref{fig:ps_mem} presents memory utilization characteristics for both Parameter Server strategies. The analysis indicates that memory utilization remains below 50\% of available GPU memory across all experimental configurations for both synchronous and asynchronous implementations. Synchronous Parameter Server configurations generally demonstrate more stable memory usage patterns across multiple GPUs, exhibiting consistent memory allocation and deallocation cycles.

Asynchronous Parameter Server implementations show slightly lower but more evenly distributed memory usage patterns. However, for memory-intensive architectures such as DenseNet, synchronous configurations provide smoother and more predictable memory performance, reducing the likelihood of memory fragmentation and allocation inefficiencies.

\subsubsection{Performance Trade-off Analysis}

The experimental results reveal a fundamental trade-off between training efficiency and model accuracy when comparing synchronous and asynchronous Parameter Server strategies. While asynchronous configurations achieve consistent improvements in training time performance (ranging from 0.68\% to 28\% across different models and configurations), these gains are accompanied by substantial reductions in final model accuracy.

The accuracy penalties associated with asynchronous training (ranging from 4.2 to 17.9 percentage points) can be attributed to parameter staleness effects, where workers compute gradients using potentially outdated parameter versions. This staleness introduces inconsistencies in the optimization process, leading to suboptimal convergence characteristics and reduced final model performance. The additional accuracy degradation observed when scaling to more GPUs is consistent with fixed hyperparameters under changed step dynamics and increased communication contention, both of which are typically addressed via scale-aware tuning in applied distributed training.

The choice between synchronous and asynchronous Parameter Server configurations should be informed by the specific requirements and constraints of the training scenario. Synchronous configurations remain optimal for applications where model accuracy is paramount and training time constraints are less critical. Conversely, asynchronous configurations may be suitable for scenarios where rapid model iteration is prioritized over achieving optimal accuracy, or where system elasticity and tolerance to stragglers outweigh the need for peak convergence quality.

These findings highlight the importance of considering both computational efficiency and optimization quality when selecting distributed training strategies, particularly in Parameter Server architectures where the synchronization paradigm directly impacts both training dynamics and final model performance.

\subsection{Experiment 3: Modified Asynchronous Parameter Server for Enhanced Accuracy}

The third experimental evaluation investigates optimization strategies for asynchronous Parameter Server configurations aimed at mitigating accuracy degradation while preserving computational efficiency benefits. This analysis examines the implementation of modified asynchronous Parameter Server approaches that incorporate enhanced communication frequency and refined gradient update mechanisms to address the accuracy limitations observed in standard asynchronous configurations.

\begin{table}[htbp]
\caption{Training Time (in seconds) for Synchronous and Modified Asynchronous Strategies}\label{tab:modified_async_ps_restructured}
\begin{tabular}{@{}lllll@{}}
\toprule
\textbf{Strategy} & \textbf{Model} & \textbf{1 GPU} & \textbf{2 GPUs} & \textbf{4 GPUs} \\
\midrule
\multirow{3}{*}{Synchronous}
& DenseNet   & 509.88 & 231.04 & 137.34 \\
& ResNet50   & 468.31 & 246.54 & 129.24 \\
& VGG16      & 891.52 & 545.61 & 247.66 \\
\midrule
\multirow{3}{*}{Modified Async}
& DenseNet   & $>$3000 & 246.54 & 250.84 \\
& ResNet50   & $>$3000 & $>$3000 & 206.64 \\
& VGG16      & $>$3000 & $>$3000 & 417.22 \\
\botrule
\end{tabular}
\end{table}

\begin{table}[htbp]
\caption{Accuracy Improvement: Standard vs. Modified Asynchronous (4-GPU Configuration)}\label{tab:accuracy_improvement}
\begin{tabular}{@{}llll@{}}
\toprule
\textbf{Model} & \textbf{Standard Async} & \textbf{Modified Async} & \textbf{Improvement} \\
\midrule
DenseNet   & 28.4\% & 44.1\% & +15.7\% \\
ResNet50   & 34.7\% & 39.5\% & +4.8\% \\
VGG16      & 42.4\% & 46.8\% & +4.4\% \\
\botrule
\end{tabular}
\end{table}

\subsubsection{Modification Strategy and Implementation}

The modified asynchronous Parameter Server implementation incorporates several optimization techniques designed to reduce parameter staleness effects while maintaining asynchronous operation characteristics. These modifications include increased communication frequency between worker nodes and parameter servers, implementation of bounded staleness mechanisms to limit the temporal divergence of parameter versions, and adaptive gradient update strategies that account for the age of gradient information during parameter updates.

The enhanced communication protocol requires workers to synchronize with parameter servers more frequently than standard asynchronous implementations, reducing the likelihood of operating on significantly outdated parameter versions. Additionally, gradient update mechanisms incorporate temporal weighting factors that adjust the influence of gradient updates based on their staleness, enabling more stable convergence characteristics while preserving asynchronous operational benefits.

\subsubsection{Training Time Performance Analysis}

Table~\ref{tab:modified_async_ps_restructured} presents training time measurements for both synchronous and modified asynchronous Parameter Server configurations across all experimental setups. The results demonstrate that the accuracy-enhancing modifications introduce substantial computational overhead, resulting in significantly increased training times compared to both standard asynchronous and synchronous implementations.

For the VGG16 architecture in the 4-GPU configuration, modified asynchronous Parameter Server requires 417.22 seconds for training completion compared to 247.66 seconds for the synchronous implementation, representing a 68.5\% increase in training time. This substantial performance penalty contradicts the fundamental objective of asynchronous training, which seeks to improve computational efficiency through reduced synchronization overhead.

The modified asynchronous approach exhibits even more pronounced performance degradation for DenseNet and ResNet50 architectures. In several configurations, particularly single-GPU and 2-GPU setups, the modified asynchronous implementation exceeds the established time limit threshold of 3000 seconds, indicating practical infeasibility for deployment scenarios with temporal constraints.

The scaling analysis reveals that modified asynchronous configurations demonstrate poor scaling characteristics, with training times increasing or remaining elevated when transitioning from 2-GPU to 4-GPU configurations. For DenseNet, the 4-GPU modified asynchronous configuration requires 250.84 seconds compared to 246.54 seconds for the 2-GPU setup, indicating minimal or negative scaling benefits.

\subsubsection{Computational Efficiency Trade-offs}

The experimental results reveal that while the implemented modifications successfully address accuracy limitations inherent in standard asynchronous Parameter Server configurations (as evidenced by the significant improvements in Table~\ref{tab:accuracy_improvement}), they introduce prohibitive computational overhead that negates the primary advantages of asynchronous distributed training. The enhanced communication frequency and sophisticated gradient update mechanisms require additional computational resources and network bandwidth, resulting in training times that exceed even synchronous implementations.

The frequent timeout occurrences in single-GPU and 2-GPU configurations indicate that the modification overhead scales poorly with reduced parallelization, suggesting that the enhanced communication protocols become increasingly inefficient when distributed across fewer computational nodes. This scaling behavior limits the practical applicability of modified asynchronous approaches to scenarios with substantial computational resources.

\subsubsection{Practical Implementation Considerations}

The analysis demonstrates that achieving improved accuracy in asynchronous Parameter Server configurations through enhanced communication and gradient update mechanisms introduces fundamental trade-offs that may compromise the practical viability of such approaches. The substantial increase in training times observed across all model architectures suggests that the computational overhead associated with staleness mitigation strategies can exceed the benefits derived from asynchronous operation.

The timeout behavior observed in multiple experimental configurations indicates that further optimization is required to develop asynchronous Parameter Server variants that achieve acceptable accuracy without sacrificing computational efficiency. Potential optimization directions include selective synchronization strategies, adaptive communication protocols that dynamically adjust based on convergence characteristics, and hybrid approaches that combine asynchronous and synchronous elements based on training phase requirements.

\subsubsection{Performance Optimization Requirements}

The experimental findings suggest that effective enhancement of asynchronous Parameter Server accuracy requires sophisticated optimization techniques that minimize communication overhead while maintaining staleness control mechanisms. The current implementation demonstrates that naive increases in communication frequency and gradient update complexity result in performance characteristics that are inferior to both standard asynchronous and synchronous approaches.

Future research directions should focus on developing intelligent adaptation mechanisms that balance accuracy improvement with computational efficiency, potentially incorporating machine learning techniques to optimize communication patterns and gradient update strategies based on real-time training dynamics. Additionally, investigation of hardware-specific optimizations and advanced network protocols may enable more efficient implementation of enhanced asynchronous Parameter Server configurations.

These results emphasize the complexity of optimizing distributed training strategies and highlight the importance of comprehensive performance evaluation when developing novel approaches that attempt to address fundamental trade-offs between computational efficiency and optimization quality in distributed machine learning systems.

\begin{table}[htbp]
\caption{Comparison of DDP, FSDP, and Parameter Server Models}\label{tab:model_comparison}
\begin{tabular*}{\textwidth}{@{\extracolsep\fill}p{2cm}p{3cm}p{3.5cm}p{4cm}}
\toprule
\textbf{Feature} & \textbf{DDP (Distributed Data Parallelism)} & \textbf{FSDP (Fully Sharded Data Parallelism)} & \textbf{PS (Parameter Server)} \\
\midrule
\textbf{Model Replication} & Full model replicated on each GPU & Model is sharded across GPUs to reduce memory usage & Model parameters are stored on central servers, while worker nodes perform training \\
\midrule
\textbf{Memory Usage} & High, as each GPU holds a complete copy of the model & Low, as parameters, gradients, and optimizer states are sharded across GPUs & Varies; lower in asynchronous mode due to decentralized updates, higher in synchronous mode \\
\midrule
\textbf{Communication Overhead} & Moderate, requires synchronization of gradients between GPUs after each batch & High, due to synchronization of sharded model parameters and gradients across GPUs & Low in asynchronous mode, but higher in synchronous mode due to global updates \\
\midrule
\textbf{Scalability} & Good for moderate GPU counts & Highly scalable for large models, as memory is distributed across GPUs & Asynchronous mode scales well with many worker nodes, while synchronous mode faces synchronization bottlenecks \\
\midrule
\textbf{Training Time} & Fast, as each GPU processes the full model in parallel and synchronizes gradients & Slower, as extra time is required for gathering sharded parameters and synchronizing them & Fast in asynchronous mode, but slower in synchronous mode due to global synchronization \\
\midrule
\textbf{Model Consistency} & High, as gradients are synchronized across all GPUs for consistent updates & High, but requires careful management of synchronization & Low in asynchronous mode due to independent updates; high in synchronous mode with global consistency \\
\midrule
\textbf{Parallelism Strategy} & Data parallelism, where each GPU processes a different subset of the data & Data and model sharding, where each GPU processes a portion of the model and its data & Parameter server architecture, where worker nodes handle different data and send updates to central servers \\
\midrule
\textbf{Distribution Method} & Each GPU holds a replica of the model and processes different data batches & The model is divided (sharded) across GPUs, reducing per-GPU memory load & Worker nodes compute gradients and update central server parameters asynchronously or synchronously \\
\midrule
\textbf{Complexity} & Moderate, easy to implement and widely used & High, as sharding introduces complexity and requires advanced synchronization & Moderate to high, depending on whether asynchronous or synchronous mode is used \\
\botrule
\end{tabular*}
\end{table}

\section{Discussion}

The experimental evaluation presented in this study provides comprehensive insights into the performance characteristics and practical trade-offs inherent in contemporary distributed deep learning strategies. The findings reveal fundamental design tensions between computational efficiency, memory utilization, and model accuracy that have significant implications for large-scale deep learning deployment strategies and the evolution of distributed training architectures.

\subsection{Implications of DDP versus FSDP Performance Characteristics}

The comparative analysis of Distributed Data Parallel and Fully Sharded Data Parallel strategies reveals a critical performance dichotomy that reflects broader challenges in distributed system design. The consistent 2-3× training time advantage demonstrated by DDP across all experimental configurations can be attributed to the computational simplicity of complete model replication, which eliminates the need for complex parameter gathering and scattering operations during forward and backward propagation phases. This performance advantage aligns with theoretical expectations for communication-optimal distributed algorithms, where minimizing the frequency of collective operations directly translates to reduced training time overhead.

However, the 4-6× memory efficiency improvement achieved by FSDP represents a paradigm shift in distributed training capability, enabling model training scenarios that would be impossible under traditional data-parallel approaches. This memory efficiency advantage becomes increasingly critical as model architectures continue to grow in complexity and parameter count, potentially reaching the limits of individual GPU memory capacity. The experimental results suggest that FSDP's parameter sharding mechanism effectively distributes memory load across participating devices while maintaining computational correctness, albeit at the expense of increased communication overhead.

The implications of these findings extend beyond immediate performance considerations to fundamental questions about resource allocation and system design in distributed deep learning environments. Organizations with abundant memory resources may prioritize DDP's training time efficiency, while those facing memory constraints or seeking to train larger models may find FSDP's memory efficiency advantages compelling despite the associated performance penalties. This trade-off relationship suggests that optimal distributed training strategy selection requires careful consideration of hardware constraints, model characteristics, and performance objectives.

\subsection{Parameter Server Architecture Trade-offs and Convergence Dynamics}

The evaluation of synchronous versus asynchronous Parameter Server configurations reveals complex interactions between system design choices and optimization quality that have profound implications for distributed training effectiveness. The consistent training time improvements achieved by asynchronous configurations (ranging from 0.68\% to 28\% across different scenarios) demonstrate the computational benefits of eliminating synchronization barriers and enabling independent worker operation. This performance advantage aligns with theoretical predictions for asynchronous distributed optimization, where eliminating coordination overhead should directly translate to improved system throughput.

However, the substantial accuracy penalties associated with asynchronous operation (ranging from 4.2 to 17.9 percentage points) highlight the fundamental tension between computational efficiency and optimization quality in distributed machine learning systems. These accuracy reductions can be attributed to parameter staleness effects, where workers operate on potentially outdated parameter versions, introducing noise into the gradient estimation process and leading to suboptimal convergence trajectories. This phenomenon reflects broader challenges in distributed optimization theory, where the trade-off between communication frequency and convergence quality represents a fundamental design constraint.

The observed accuracy degradation across both synchronous and asynchronous configurations as GPU count increases suggests that parallelization introduces inherent coordination challenges that affect optimization dynamics. This finding aligns with recent theoretical work on the limits of parallel optimization, indicating that increased parallelization may introduce difficulties in maintaining convergence quality, often exacerbated by the implicit increase in global batch size which requires careful hyperparameter scaling. The implications of these findings suggest that Parameter Server architectures may be most suitable for applications where rapid iteration and approximate solutions are acceptable, rather than scenarios requiring optimal model performance.

\subsection{Challenges in Asynchronous Optimization Enhancement}

The investigation of modified asynchronous Parameter Server strategies reveals the complexity of addressing fundamental limitations in distributed optimization systems. The implemented modifications, including enhanced communication frequency and temporal gradient weighting, represent principled approaches to mitigating parameter staleness effects while preserving asynchronous operational benefits. However, the resulting performance characteristics demonstrate that such modifications can introduce computational overhead that exceeds the benefits of the underlying asynchronous approach.

The frequent timeout occurrences and poor scaling characteristics observed in modified asynchronous configurations indicate that naive approaches to staleness mitigation may fundamentally compromise system efficiency. This outcome suggests that effective enhancement of asynchronous distributed training requires sophisticated algorithmic innovations that balance staleness control with communication efficiency. The experimental results align with recent theoretical work on bounded staleness algorithms, which indicate that effective asynchronous optimization enhancement requires careful consideration of communication patterns, gradient aging effects, and adaptive synchronization mechanisms.

These findings have broader implications for the development of next-generation distributed training systems, suggesting that hybrid approaches combining synchronous and asynchronous elements may be necessary to achieve optimal performance across diverse deployment scenarios. The results emphasize the importance of comprehensive performance evaluation when developing novel distributed training approaches, as seemingly reasonable modifications can introduce unexpected performance penalties that compromise practical viability.

\subsection{Scalability and Resource Utilization Patterns}

The experimental evaluation reveals consistent scaling patterns across all evaluated distributed training strategies, with near-linear performance improvements when increasing GPU count from 1 to 4 devices. This scaling behavior suggests that the evaluated strategies effectively utilize additional computational resources, although the absolute performance levels vary significantly between approaches. The consistent scaling characteristics indicate that the performance differences observed between strategies stem from algorithmic and communication design choices rather than fundamental limitations in parallel processing capability.

The resource utilization analysis reveals that GPU utilization patterns remain comparable across different strategies, suggesting that computational efficiency differences arise from communication and synchronization overhead rather than computational load imbalance. This finding has important implications for distributed system design, indicating that optimization efforts should focus on communication protocol enhancement and synchronization algorithm improvement rather than computational load balancing mechanisms.

The memory utilization characteristics demonstrate significant variation between strategies, with FSDP achieving substantial memory efficiency improvements compared to DDP approaches. These memory efficiency differences have critical implications for system scalability, as they directly impact the maximum model size that can be trained on available hardware. The experimental results suggest that memory-efficient distributed training strategies may enable training of significantly larger models within existing hardware constraints, potentially accelerating progress in large-scale deep learning research.

\subsection{Addressing Gang Scheduling Dependencies}

The experimental findings provide important insights into the original research motivation regarding gang scheduling dependencies in distributed deep learning systems. The evaluation demonstrates that different distributed training strategies exhibit varying degrees of synchronization requirements, with implications for resource allocation flexibility and system utilization efficiency.

The DDP approach, while achieving superior training time performance, requires strict synchronization across all participating devices for gradient aggregation and parameter updates. This synchronization requirement directly reinforces gang scheduling dependencies, as all workers must complete their computational phases before proceeding to subsequent training iterations. The FSDP strategy introduces additional synchronization complexity through parameter gathering and scattering operations, potentially exacerbating gang scheduling constraints despite offering memory efficiency advantages.

The Parameter Server architecture presents more promising opportunities for reducing gang scheduling dependencies, particularly in asynchronous configurations where workers can operate independently without strict synchronization barriers. However, the observed accuracy penalties associated with asynchronous operation suggest that reducing gang scheduling dependencies may come at the expense of optimization quality, creating a fundamental trade-off between system flexibility and model performance.

These findings suggest that addressing gang scheduling dependencies in distributed deep learning systems requires careful consideration of performance trade-offs and application requirements. While asynchronous approaches offer potential solutions for reducing synchronization constraints, their practical effectiveness depends on the specific performance objectives and accuracy requirements of the training scenario.

\subsection{Broader Implications for Distributed Deep Learning}

The experimental results contribute to broader understanding of distributed deep learning system design and optimization, revealing fundamental trade-offs that inform both theoretical research and practical deployment decisions. The findings demonstrate that no single distributed training strategy provides optimal performance across all evaluation dimensions, suggesting that effective distributed deep learning systems may require adaptive approaches that dynamically select optimization strategies based on current system conditions and performance objectives.

The consistent trade-offs observed between training time, memory utilization, and model accuracy suggest that distributed deep learning system design involves inherent compromises that cannot be eliminated through algorithmic improvements alone. These trade-offs reflect fundamental limitations in distributed computing and optimization theory, indicating that practical distributed training systems must be designed with explicit consideration of performance priorities and constraint tolerances.

The experimental findings also highlight the importance of empirical evaluation in distributed system research, as theoretical predictions about system performance may not accurately reflect real-world behavior under practical deployment conditions. The observed performance characteristics suggest that distributed training strategy selection requires careful empirical evaluation under realistic conditions rather than relying solely on theoretical performance models.

\subsection{Limitations and Future Research Directions}

The current study presents several limitations that suggest directions for future research investigation. The experimental evaluation focuses on relatively small-scale distributed training scenarios, with maximum configurations involving four GPUs. Larger-scale evaluations involving dozens or hundreds of GPUs may reveal different performance characteristics and scaling behaviors that could alter the observed trade-off relationships.

The evaluation is limited to specific model architectures and datasets, which may not fully represent the diversity of workloads encountered in practical distributed deep learning deployments. Future research should examine performance characteristics across broader ranges of model types, including transformer architectures, generative models, and specialized domain-specific architectures.

The Parameter Server implementations evaluated in this study represent classical approaches that may not reflect recent algorithmic innovations in asynchronous optimization and communication protocol design. Future research should investigate more sophisticated Parameter Server variants that incorporate advanced staleness mitigation techniques, adaptive communication protocols, and hybrid synchronization mechanisms.

Additionally, the current evaluation focuses primarily on homogeneous hardware configurations, while practical distributed deep learning deployments often involve heterogeneous hardware with varying computational capabilities and network characteristics. Future research should examine distributed training strategy performance under heterogeneous conditions and investigate adaptive approaches that accommodate hardware diversity.

The findings presented in this study establish a foundation for future research into adaptive distributed training systems that dynamically optimize strategy selection based on real-time performance characteristics and changing system conditions. Such adaptive approaches may enable more effective utilization of distributed computing resources while maintaining acceptable performance across diverse training scenarios and deployment environments.

\section{Conclusion}

This study presented a structured evaluation of three widely used distributed training strategies: Distributed Data Parallel (DDP), Fully Sharded Data Parallel (FSDP), and the Parameter Server (PS) architecture. The goal was to examine the practical differences in training time, memory usage, scalability, and convergence across multiple deep learning models and GPU configurations under consistent experimental conditions.

The results show that DDP consistently achieves the fastest training times, especially for compute-heavy models running on memory-rich clusters. However, because DDP requires full model replication on each GPU, it consumes significantly more memory. This limits its usefulness when training very large networks. FSDP offers a clear advantage, demonstrating a 4--6$\times$ reduction in per-GPU memory usage, which allows larger models to be trained on the same hardware. This comes at the cost of longer training times (approximately 2--3$\times$ slower than DDP) due to increased communication. Even with this overhead, FSDP scales well and is suitable for memory-constrained environments.

The Parameter Server strategy adds flexibility, particularly in asynchronous configurations where synchronization is relaxed. This improves training speed but leads to reduced accuracy caused by parameter staleness. Synchronous PS avoids this issue and provides more stable convergence but takes longer to train. We also tested a modified version of asynchronous PS with more frequent communication and staleness-aware updates. While it substantially recovered accuracy (e.g., +15.7\% for DenseNet), it introduced considerable overhead and often took even longer than synchronous PS.

Each strategy presents its own set of trade-offs. DDP is best when training speed is the priority and memory is not a concern. FSDP is ideal when memory efficiency matters more than raw training speed. PS strategies are still useful in heterogeneous environments or when elasticity is important, though care must be taken to manage the accuracy impact of asynchronous updates.

These findings offer practical guidance for selecting distributed training methods based on system constraints and training goals. Future work should explore hybrid strategies that combine elements of these approaches and investigate performance at larger scales using a broader range of models and datasets.

\section*{Declarations}

\begin{itemize}
    \item \textbf{Funding:} The author declares that no funds, grants, or other support were received during the preparation of this manuscript.
    
    \item \textbf{Conflict of interest:} The author has no relevant financial or non-financial interests to disclose.
    
    \item \textbf{Ethics approval and consent to participate:} Not applicable. This research focuses on machine learning algorithms using publicly available datasets (CIFAR-10, MNIST) and does not involve human participants or animals.
    
    \item \textbf{Consent for publication:} Not applicable.
    
    \item \textbf{Data availability:} The datasets analysed during the current study are publicly available. The CIFAR-10 dataset is available at \url{https://www.cs.toronto.edu/~kriz/cifar.html} and the MNIST dataset is available at \url{http://yann.lecun.com/exdb/mnist/}.
    
    \item \textbf{Materials availability:} Not applicable.
    
    \item \textbf{Code availability:} The code used to generate the results in this study is available from the corresponding author upon reasonable request.
    
\end{itemize}

\bibliography{sn-bibliography}

\end{document}